\documentclass[12pt]{amsart}
\usepackage[latin9]{inputenc}
\usepackage{geometry}
\usepackage{color}
\usepackage{float}
\usepackage{units}
\usepackage{mathrsfs}
\usepackage{amsbsy}
\usepackage{amstext}
\usepackage{amsthm}
\usepackage{amssymb}
\usepackage{graphicx}
\usepackage{wasysym}
\PassOptionsToPackage{normalem}{ulem}
\usepackage{ulem}
\usepackage[unicode=true,
 bookmarks=false,
 breaklinks=false,pdfborder={0 0 1},backref=page,colorlinks=true]
 {hyperref}

\makeatletter

\numberwithin{equation}{section}
\numberwithin{figure}{section}
\theoremstyle{plain}
\newtheorem{thm}{\protect\theoremname}[section]
\theoremstyle{definition}
\newtheorem{defn}[thm]{\protect\definitionname}
\theoremstyle{remark}
\newtheorem{rem}[thm]{\protect\remarkname}
\theoremstyle{plain}
\newtheorem{prop}[thm]{\protect\propositionname}
\theoremstyle{plain}
\newtheorem{lem}[thm]{\protect\lemmaname}

\@ifundefined{date}{}{\date{}}

\usepackage{hyperref}
\usepackage{amsthm}
\usepackage{amscd}
\usepackage{comment}
\usepackage{mathrsfs}
\usepackage{setspace}
\usepackage{color}

\newcommand{\ud}{\mathrm{d}}

\DeclareMathOperator{\Or}{O}

\DeclareMathOperator{\Q}{Q}

\newcommand{\mc}{\mathcal}
\newcommand{\mf}{\mathfrak}

\newcommand{\rz}{\mathbb R}

\newcommand{\bs}{\boldsymbol}

\makeatother

\providecommand{\definitionname}{Definition}
\providecommand{\lemmaname}{Lemma}
\providecommand{\propositionname}{Proposition}
\providecommand{\remarkname}{Remark}
\providecommand{\theoremname}{Theorem}

\begin{document}
\global\long\def\Z{\mathbb{Z}}%

\global\long\def\E{\mathbb{E}}%

\global\long\def\R{\mathbb{R}}%

\global\long\def\rz{\mathbb{R}}%

\global\long\def\C{\mathbb{C}}%

\global\long\def\N{\mathbb{N}}%

\global\long\def\Q{\mathbb{Q}}%

\global\long\def\ud{\mathrm{d}}%

\global\long\def\Cr{\mathscr{C}\left(f\right)}%

\global\long\def\Min{\mathscr{M}_{-}\left(f\right)}%

\global\long\def\Max{\mathscr{M}_{+}\left(f\right)}%

\global\long\def\Sd{\mathscr{S}\left(f\right)}%

\global\long\def\Xt{\mathscr{X}\left(f\right)}%

\global\long\def\Nd{\mathcal{N}\left(f\right)}%

\global\long\def\Or{\mathrm{O}}%

\global\long\def\hess{\mathrm{Hess}}%

\global\long\def\mf#1{\mathfrak{#1}}%

\global\long\def\mc#1{\mathcal{#1}}%

\global\long\def\bs#1{\boldsymbol{#1}}%

\global\long\def\qform{q}%

\global\long\def\dom#1{\mathrm{Dom}(#1)}%

\global\long\def\spec#1{\mathrm{Spec}(#1)}%

\global\long\def\supp{\mathrm{\mathop{supp}}}%

\global\long\def\stardom{\Omega_{a,b}^{\mathrm{star}}}%

\global\long\def\bdry{\gamma_{a,b}}%

\global\long\def\lensdom{\Omega_{a,b}^{\mathrm{lens}}}%

\global\long\def\quarterdom{\Lambda_{a,b}}%

\global\long\def\lap{\Delta}%

\global\long\def\lapstar{\Delta_{a,b}}%

\global\long\def\laph{\Delta_{a,b}^{h}}%

\global\long\def\lapv{\Delta_{a,b}^{v}}%

\global\long\def\sector{S_{R}}%

\global\long\def\lapsec{\Delta_{\alpha,R}}%

\title{The Spectral Position of Neumann domains on the torus}
\author{Ram Band$^{1}$, Sebastian K Egger$^{1}$, Alexander J Taylor$^{2}$}
\address{$^{1}${\small{}Department of Mathematics, Technion--Israel Institute
of Technology, Haifa 32000, Israel.}}
\address{$^{2}$This work was carried out while Alexander Taylor was employed
by the University of Bristol.\\
H H Wills Physics Laboratory, University of Bristol, Tyndall Avenue,
Bristol BS8 1TL, United Kingdom.}
\begin{abstract}
Neumann domains of Laplacian eigenfunctions form a natural counterpart
of nodal domains. The restriction of an eigenfunction to one of its
nodal domains is the \emph{first} Dirichlet eigenfunction of that
domain. This simple observation is fundamental in many works on nodal
domains. We consider a similar property for Neumann domains. Namely,
given a Laplacian eigenfunction $f$ and its Neumann domain $\Omega$,
what is the position of $\left.f\right|_{\Omega}$ in the Neumann
spectrum of $\Omega$? 

The current paper treats this spectral position problem on the two-dimensional
torus. We fully solve it for separable eigenfunctions on the torus
and complement our analytic solution with numerics for random waves
on the torus. These results answer questions from \cite{Zel_sdg13,BanFaj_ahp16}
and arouse new ones.
\end{abstract}

\maketitle

\section{Introduction}

\subsection{Neumann domains}

\noindent Let $(M,\thinspace g)$ be a two-dimensional, connected,
orientable and closed Riemannian manifold. We denote by $-\Delta$
the (negative) self-adjoint Laplace-Beltrami operator. Its spectrum
is purely discrete since $M$ is compact. We order the eigenvalues
$\{\lambda_{n}\}_{n=0}^{\infty}$ increasingly, $0=\lambda_{0}<\lambda_{1}\leq\lambda_{2}\leq\ldots$,
and denote a corresponding complete system of orthonormal eigenfunctions
by $\{f_{n}\}_{n=0}^{\infty}$, so that we have
\begin{equation}
-\Delta f_{n}=\lambda_{n}f_{n}.
\end{equation}

\noindent Let $f$ be an eigenfunction (we suppress its position,
for brevity). We introduce a flow along the gradient vector field,
$\nabla f$,
\begin{equation}
\begin{aligned} & \varphi:\mathbb{R}\times\,M\rightarrow M,\\
 & \partial_{t}\varphi(t,\,\bs x)=-\nabla f\big|_{\varphi(t,\,\bs x)},\\
 & \varphi(0,\,\bs x)=\bs x.
\end{aligned}
\label{eq:flow}
\end{equation}

\noindent The set of critical points of $f$ is denoted by $\Cr$;
the sets of saddle points and extrema of $f$ are denoted by $\Sd$
and $\Xt$; the sets of minima and maxima of $f$ are denoted by $\Min$
and $\Max$, respectively.

\noindent For a critical point $\bs x\in\Cr$, we define its stable
and unstable manifolds by
\begin{equation}
\begin{aligned}W^{s}(\bs x) & :=\{\bs y\in M~~:~~\lim_{t\rightarrow\infty}\varphi(t,\,\bs y)=\bs x\}\quad\mbox{ and }\\
W^{u}(\bs x): & =\{\bs y\in M~~:~~\lim_{t\rightarrow-\infty}\varphi(t,\,\bs y)=\bs x\},
\end{aligned}
\label{eq:Stable-Unstable-Def}
\end{equation}
respectively.

\noindent We assume in the following that the eigenfunctions $f$
are Morse functions, i.e.,\, the determinant of the Hessian does
not vanish at critical points. We call such an $f$ a \emph{Morse
eigenfunction}. Eigenfunctions are generically Morse, as was shown
in \cite{Albert_thesis72,Uhl_ajm76}.
\begin{defn}
\noindent \cite{BanFaj_ahp16}\label{def:Neumann-Domains-and-Lines}
Let $f$ be a Morse function.
\begin{enumerate}
\item Let $\bs p\in\Min,\,\,\bs q\in\Max$, such that $W^{s}\left(\bs p\right)\cap W^{u}\left(\bs q\right)\neq\emptyset$.
Each of the connected components of $W^{s}\left(\bs p\right)\cap W^{u}\left(\bs q\right)$
is called a \emph{Neumann domain} of $f$.
\item The \emph{Neumann line set} of $f$ is 
\begin{equation}
\Nd:=\overline{\bigcup_{{\bs r\in\Sd}}W^{s}(\bs r)\cup W^{u}(\bs r)}.\label{eq:Neumann-line-set}
\end{equation}
\end{enumerate}
\end{defn}

\begin{figure}[h]
\centering{}\includegraphics[width=1\textwidth]{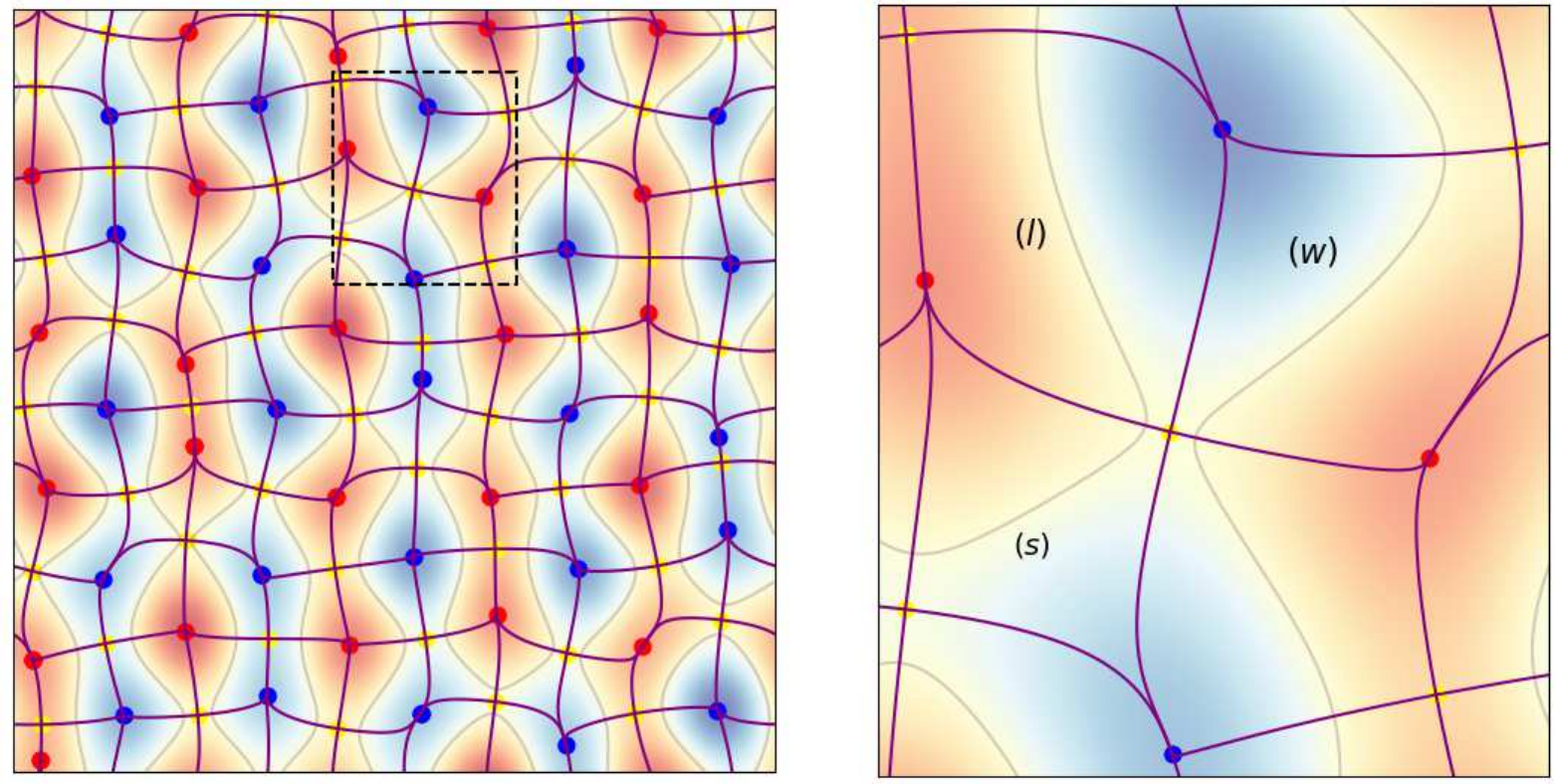}
\caption{\uline{Left}: An eigenfunction corresponding to the eigenvalue
$\lambda=25$ of the flat torus whose fundamental domain is $[0,2\pi]\times[0,2\pi]$.
Red (blue) colors indicate positive (negative) values of the eigenfunction.
Red (blue) circles mark maxima (minima) and yellow circles mark saddle
points. The nodal set is drawn in grey and the Neumann line set in
purple. The Neumann domains are the domains bounded by the Neumann
line set.\protect \\
\uline{Right}: A magnification of the marked square from the left
figure. Three Neumann domains are marked by (s), (l) and (w) (which
stand for star, lens and wedge) according to the three distinguished
Neumann domain types described in Section \ref{subsec:Numerics-random-waves}.}
\label{fig:Neumann-Domains-Torus-Basic} 
\end{figure}

Note that the definition above may be applied to any Morse function
and not necessarily to eigenfunctions. Nevertheless, in this paper
we are interested to study the Neumann domains and Neumann lines of
Laplacian eigenfunctions.\\
It follows from basic Morse theory that Neumann domains are two-dimensional
subsets of $M$, whereas the Neumann line set is a union of one dimensional
curves on $M$. As an example, see Figure \ref{fig:Neumann-Domains-Torus-Basic}
which shows an eigenfunction of the flat torus with its partition
to Neumann domains.

\subsection{Spectral position \label{subsec:Spectral-position}}

Let $f$ be an eigenfunction corresponding to an eigenvalue $\lambda$
and let $\Omega$ be a Neumann domain of $f$. The boundary, $\partial\Omega$,
consists of Neumann lines, which are particular gradient flow lines.
As the gradient $\nabla f$ is tangential to the Neumann lines we
get that $\left.\partial_{\nu}f\right|_{\partial\Omega}:=\left.\hat{\nu}\cdot\nabla f\right|_{\partial\Omega}=0$,
where $\hat{\nu}$ is normal to $\partial\Omega$. As a consequence
we have that $\left.f\right|_{\Omega}$ is a Neumann eigenfunction
of $\Omega$ and corresponds to the eigenvalue $\lambda$, which is
the reason behind the name \emph{Neumann domains}. The proof that
$\left.f\right|_{\Omega}$ is a Neumann eigenfunction of a general
Neumann domain $\Omega$, goes beyond the classical Sobolev embedding
theorems\footnote{The reason for this is that the boundary of a general Neumann domain
might include a cusp and in general we do not have an explicit expression
of the cusp.} and appears in \cite{BanCoxEgg_prep19}. In the current paper we
only supply a proof suited for the particular Neumann domains treated
here (Proposition \ref{prop:domain_of_star_Laplacian},(\ref{enu:domain_of_star_Laplacian-2})
and Remark \ref{rem:Lens_domain_spectral_prop}).

Following the discussion above we get that $\lambda$ is a Neumann
eigenvalue of $\Omega$. It is natural to ask what is the position
of this $\lambda$ in the Neumann spectrum of $\Omega$.
\begin{defn}
\label{def:Spectral-Position} Let $f$ be a Morse eigenfunction of
an eigenvalue $\lambda$ and let $\Omega$ be a Neumann domain of
$f$. We define the spectral position of $\Omega$ as the position
of $\lambda$ in the Neumann spectrum of $\Omega$. It is explicitly
given by 
\begin{equation}
N_{\Omega}(\lambda):=\left|\left\{ \lambda_{n}\in\mathrm{Spec}(\Omega)~:~\lambda_{n}<\lambda\right\} \right|,\label{eq:Spectral-Position}
\end{equation}
where $\mathrm{Spec}(\Omega):=\{\lambda_{n}\}_{n=0}^{\infty}$ is
the pure point part of the Neumann spectrum of $\Omega$, containing
multiple appearances of degenerate eigenvalues and including $\lambda_{0}=0$.
\end{defn}

\begin{rem}
\label{rem:After-def-spectral-position}~
\begin{enumerate}
\item If $\lambda$ is a multiple eigenvalue of $\Omega$, then by this
definition the spectral position is the lowest position of $\lambda$
in the spectrum.
\item The spectral position is positive for any Neumann domain, i.e., $N_{\Omega}(\lambda)>0$.
Indeed, $N_{\Omega}(\lambda)=0$ is possible only for $\lambda=0$,
but the zero eigenvalue corresponds to the constant eigenfunction
which does not have Neumann domains at all.
\end{enumerate}
\end{rem}

For comparison, we mention what is the spectral position for \emph{nodal}
domains. Consider a nodal domain $\Xi$ of some eigenfunction $f$
corresponding to an eigenvalue $\lambda$. It is known that $\left.f\right|_{\Xi}$
is the first eigenfunction (aka ground-state) of $\Xi$ with Dirichlet
boundary conditions \cite{Courant23}. Namely, $\lambda$ is the lowest
eigenvalue in the Dirichlet spectrum of $\Xi$, or if adopting the
notation (\ref{def:Spectral-Position}) for the Dirichlet spectrum
we get $N_{\Xi}(\lambda)=0$. This observation is fundamental in many
results concerning nodal domains and their counting. 

The purpose of this paper is to study the spectral positions of Neumann
domains. The general problem is quite involved (comparing to the easy
answer for nodal domains, as mentioned above) and in this paper we
concentrate on investigating this problem for the two-dimensional
flat torus.

A qualitative feeling on the value of $N_{\Omega}(\lambda)$ might
be given by \cite[Theorem 1.4]{BanFaj_ahp16} (see also \cite[Theorem 3.2]{AloBanBerEgg_Neumann}).
It is shown there that the ``topography'' of $f|_{\Omega}$ cannot
be too complex; its domain, $\Omega$, is simply connected; $f|_{\Omega}$
has no critical points in the interior of $\Omega$; and its zero
set is merely a single simple non-intersecting curve. These observations
suggest that $f|_{\Omega}$ might not lie too high in the spectrum
of $\Omega$. Such a belief appears also in \cite{Zel_sdg13}, where
it is written that possibly, the spectral position of Neumann domains
'often' equals one, just as in the case of nodal domains. Our task
is to study the possible values of $N_{\Omega}(\lambda)$ and to find
out to what extent $\lambda$ is indeed the first non trivial eigenvalue
of $\Omega$ ($N_{\Omega}(\lambda)=1$).

\subsection{Torus eigenfunctions}

We consider the flat torus with fundamental domain $\R^{2}/\Z^{2}$
equipped with the Laplace operator. The eigenvalues are 
\begin{align}
\lambda_{a,b}: & =\frac{\pi^{2}}{4}\left(\frac{1}{a^{2}}+\frac{1}{b^{2}}\right),\label{eq:Torus-eigenvalue}
\end{align}
 where
\begin{equation}
a:=\frac{1}{4m_{x}},~~b:=\frac{1}{4m_{y}},~~~\quad\textrm{for }m_{x},m_{y}\in\N.\label{eq:quantum-numbers-separable-efunc}
\end{equation}

The separable eigenfunctions may be written as
\begin{equation}
f_{a,b}(x,y)=\cos\left(\frac{\pi}{2a}x\right)\cos\left(\frac{\pi}{2b}y\right).\label{eq:Torus-eigenfunction}
\end{equation}

\begin{figure}
\centering{}\includegraphics[width=1\textwidth]{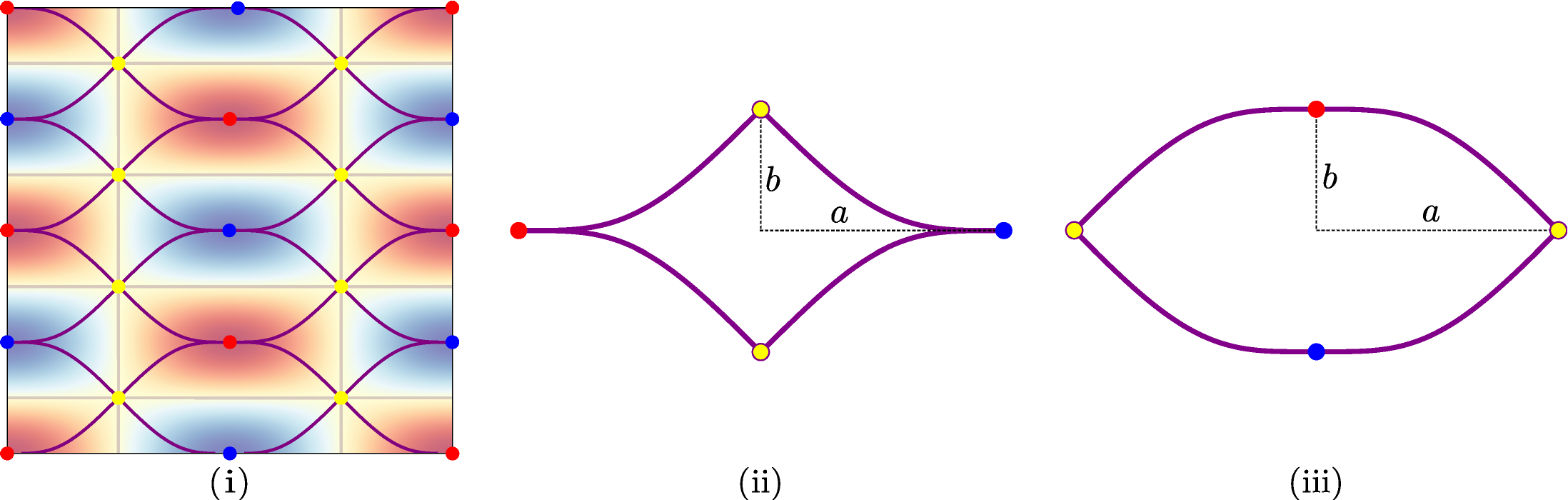}
\caption{(i): The torus eigenfunction $f(x,y)=\cos(2\pi x)\cos(4\pi y)$ with
grey lines indicating its nodal set and purple lines indicating the
Neumann line set. (ii) and (iii): the star-like and lens-like Neumann
domains of a separable eigenfunction (\ref{eq:Torus-eigenfunction}),
with the typical lengths $a,\thinspace b$ marked as dashed lines.
Saddle points are marked by yellow circles and extrema by blue and
red circles.}
\label{fig:Torus-with-star-and-lens} 
\end{figure}
Each of those eigenfunctions has two types of Neumann domains. Half
of them are of a lens shape and congruent to each other and the other
half are of a star shape and also congruent (Figure \ref{fig:Torus-with-star-and-lens}).
We denote those domains by $\stardom$ (Figure \ref{fig:Torus-with-star-and-lens},(ii))
and $\lensdom$ (Figure \ref{fig:Torus-with-star-and-lens},(iii)),
respectively. The size of those Neumann domains is determined by the
values of $a,b$ in (\ref{eq:quantum-numbers-separable-efunc}). We
wish to study the spectral position of those Neumann domains.

First, observe that the symmetry of the problem allows us to consider
only the case $b\leq a$. Second, the spectral position of either
$\stardom$ or $\lensdom$ depends only on the ratio $\frac{a}{b}$,
as rescaling both $a$ and $b$ by the same factor amounts to an appropriate
rescaling of the Neumann domain together with its eigenfunction restriction.
We then have the following.
\begin{thm}
\label{thm:Spectral-pos-separable}
\end{thm}

\begin{enumerate}
\item \label{enu:thm-spectral-pos-lens} The set of spectral positions of
the \textbf{lens}-like domains $\left\{ N_{\lensdom}\left(\lambda_{a,b}\right)\right\} _{a,b}$
is unbounded. In particular, $N_{\lensdom}\left(\lambda_{a,b}\right)\rightarrow\infty$
for $\frac{a}{b}\rightarrow\infty$.
\item \label{enu:thm-spectral-pos-star-is-one} There exists $c>1$ such
that if $\nicefrac{a}{b}>c$ then the spectral position of the \textbf{star}-like
domains equals one, i.e., $N_{\stardom}(\lambda_{a,b})=1$. In addition
$\lambda_{a,b}$ is a simple Neumann eigenvalue of $\stardom$.
\end{enumerate}
\begin{rem}
We may provide an estimate for the constant in the second part of
the theorem, which is $c\approx1.1407$. This is done in the course
of the theorem's proof (see Remark \ref{rem:Numeric_value_of_constant}).
That this constant is close to one means that the result of the theorem
is quite close to being optimal (since we always assume $a>b$). It
is interesting to find out whether the result in the theorem actually
holds for all $a>b$.
\end{rem}

\begin{rem}
In \cite[Proposition 1.7]{BanFaj_ahp16} it was proven that $\left\{ N_{\lensdom}(\lambda_{a,b})\right\} _{a,b}\cup\left\{ N_{\stardom}(\lambda_{a,b})\right\} _{a,b}$
is unbounded. The first part of Theorem \ref{thm:Spectral-pos-separable}
is a refinement of that result, showing that the unboundedness of
the spectral position is due to the lens-like Neumann domains.
\end{rem}

The first part of Theorem \ref{thm:Spectral-pos-separable} is surprising
in the light of the intuition described at the end of Section \ref{subsec:Spectral-position}.
The original expectation was that the spectral positions of Neumann
domains would be relatively low, if not even equal to one, as in the
case of nodal domains. The theorem above shows that spectral positions
may behave completely differently than expected. The second part of
the theorem somewhat revives the original intuition about spectral
positions. In essence, we show that the unbounded spectral positions
of the lens-like domains are compensated by the minimal spectral positions
of the other half of the Neumann domains - the star-like ones.

In general, the computation of spectral positions is not an easy task.
In particular, it is harder to show that the spectral position is
low (rather than high). For example, using test functions in the corresponding
quadratic form (aka Rayleigh-Ritz quotient) could only be used to
prove lower bounds on spectral positions. 

The outline of the paper is as follows. In the next section we bring
two proofs of the first part of Theorem \ref{thm:Spectral-pos-separable}.
The proof of the second part of Theorem \ref{thm:Spectral-pos-separable}
spreads over Sections \ref{sec:Spectral-analysis}, \ref{sec:Symmetry}
and \ref{sec:Rearrangement}. In Section \ref{sec:Spectral-analysis}
we provide the basic spectral properties of the Neumann eigenvalue
problem on $\stardom$. Section \ref{sec:Symmetry} presents a symmetry
reduction of this eigenvalue problem. Section \ref{sec:Rearrangement}
then complements the proof by solving some required eigenvalue comparison
problems. In Section \ref{sec:Numerics} we go beyond separable eigenfunctions
and combine a useful geometric parameter with numerical methods to
study the generic behavior of spectral positions. Some technical computations
needed for the proofs are deferred to Appendix \ref{sec:Appendix-area-calculation-of-star}.

\section{Two proofs of Theorem \ref{thm:Spectral-pos-separable}, part (\ref{enu:thm-spectral-pos-lens})\label{sec:Proof-of-Theorem-part-one}}
\begin{proof}
[First proof of Theorem  \ref{thm:Spectral-pos-separable}, (\ref{enu:thm-spectral-pos-lens})]

Assume by contradiction that there exists a constant $M$ such that
$\forall a,b~~~$ $\lambda_{a,b}\leq\lambda_{M}(\lensdom)$. By \cite{Kro_jfa92}
we have $\lambda_{M}(\lensdom)\left|\lensdom\right|\leq8\pi M$, where
$\left|\lensdom\right|$ denotes the area of $\lensdom$. Combining
this with the contradiction assumption we get
\begin{equation}
\forall a,b\quad\quad\lambda_{a,b}\left|\lensdom\right|\leq8\pi M.\label{eq:lens_lambda_upper_bound}
\end{equation}
 In addition, 
\begin{align}
\forall a,b\quad\quad\lambda_{a,b}\left|\lensdom\right| & =\frac{\pi^{2}}{4}\left(\frac{1}{a^{2}}+\frac{1}{b^{2}}\right)\left|\lensdom\right|\nonumber \\
 & =\pi^{2}\left(\frac{a}{b}+\frac{b}{a}\right)\left(\frac{1}{4ab}\left|\lensdom\right|\right)\nonumber \\
 & =\pi^{2}\left(\frac{a}{b}+\frac{b}{a}\right)\left(\frac{1}{4ab}\left(4ab-\left|\stardom\right|\right)\right)\nonumber \\
 & >\pi^{2}\left(\frac{a}{b}+\frac{b}{a}\right)-\frac{\pi^{2}}{4}\frac{2}{\pi}(j_{0})^{2}\label{eq:lower_bound_on_lens_area_times_lambda}
\end{align}
where we used that $\left|\lensdom\right|+\left|\stardom\right|=4ab$
(since the union of a quarter of $\lensdom$ and a quarter of $\stardom$
gives a rectangle $a\times b$) and the last line is a consequence
of Lemma \ref{eq:Lemma-bound_on_star_area}.

Taking the limit $\frac{a}{b}\rightarrow\infty$ in (\ref{eq:lower_bound_on_lens_area_times_lambda})
we get $\lambda_{a,b}\left|\lensdom\right|\rightarrow\infty$, which
contradicts (\ref{eq:lens_lambda_upper_bound}). Hence $\left\{ N_{\lensdom}(\lambda_{a,b})\right\} _{a,b}$
is unbounded.
\end{proof}
\newpage{}
\begin{proof}
[Second proof of Theorem  \ref{thm:Spectral-pos-separable}, (\ref{enu:thm-spectral-pos-lens})]

The lens domain $\lensdom$ is bounded within a rectangle of width
$2a$ and height $2b$ (see Figure \ref{fig:Torus-with-star-and-lens},(iii)).
When taking the limit $b\rightarrow0$ the lens domain shrinks into
a one edge graph of length $2a$. Applying results from \cite{KucZen_incol03,KucZen_jmaa01,RubSch_arma01}
(for example, we may use \cite[Theorem 3.1]{KucZen_incol03} with
$\alpha=1$) we get the eigenvalue convergence
\[
\lim_{b\rightarrow0}\lambda_{n}(\lensdom)=\left(\frac{\pi}{2a}n\right)^{2}.
\]
 Comparing this with the eigenvalue of the separable eigenfunction
\[
\lambda_{a,b}=\frac{\pi^{2}}{4}\left(\frac{1}{a^{2}}+\frac{1}{b^{2}}\right),
\]
 we see that when fixing the value of $a$ and letting $b\rightarrow0$
the spectral position of $\lambda_{a,b}$ in the spectrum of $\lensdom$
is indeed unbounded.
\end{proof}

\section{Basic spectral analysis on $\protect\stardom$ \label{sec:Spectral-analysis}}

In this section we state and prove some basic spectral properties
of the Neumann domain $\stardom$ which are needed for the proof of
Theorem \ref{thm:Spectral-pos-separable},(\ref{enu:thm-spectral-pos-star-is-one}).

The domain $\stardom$ is given by
\[
\stardom=\left\{ \left(x,y\right):~~-a<x<a,~~-\bdry(x)<y<\bdry(x)\right\} 
\]
where 
\begin{equation}
\gamma_{a,b}(x):=\frac{2b}{\pi}\arcsin\left(\left[\cos\left(\frac{\pi}{2a}x\right)\right]^{\left(\frac{a}{b}\right)^{2}}\right).\label{eq:boundary_of_star_domain}
\end{equation}

See Lemma \ref{lem:gamma_formula} where the boundary curve $\bdry$
is explicitly calculated.

In order to analyze the Neumann spectrum of $\stardom$ we need a
suitable description of the operator and the relevant quadratic form.
The domain $\stardom$ possesses a cusp (at $x=\pm a$, $y=0$), which
prohibits the standard application of the Gauss-Green identity (integration
by parts), cf. \cite{AcoArmDurLom_jmaa05} and prevents a direct characterization
of the Neumann Laplacian in terms of Neumann boundary conditions.
Instead, we use the approach which describes a semibounded self-adjoint
operator by its uniquely associated quadratic form.

We start by introducing a sesquilinear form on $\stardom$ which then
generates in a canonical way (\cite[Theorem~VIII.15]{ReedSimon_v14})
the operator $\lapstar$, which is the self-adjoint Neumann Laplacian
on $\stardom$. The form and its domain are
\begin{equation}
\begin{split}\qform[\phi,\psi] & :=\int_{\stardom}\langle\nabla\psi(x),\nabla\phi(x)\rangle_{\C^{2}}\ud x,\\
\dom{\qform} & :=W^{1,2}(\stardom),
\end{split}
\label{eq:form_domain_star}
\end{equation}
where $W^{1,2}(\stardom)$ denotes the corresponding Sobolev space
on $\stardom$. Indeed, $\qform[\cdot,\cdot]$ yields an appropriate
and well-defined Laplacian as the following proposition shows.
\begin{prop}
\label{prop:spectral-properties-of-star-domain}~ 
\begin{enumerate}
\item \label{enu:spectral-properties-of-star-domain-1} $\left(\qform,\dom{\qform}\right)$
defines a unique self-adjoint operator $\lapstar$, 
\item \label{enu:spectral-properties-of-star-domain-2} $\lapstar$ has
purely discrete spectrum.
\end{enumerate}
\end{prop}

\begin{proof}
The standard approach \cite[Theorem VIII.15]{ReedSimon_v14} to verify
part (\ref{enu:spectral-properties-of-star-domain-1}) is to realize
that the form is non-negative and to show that the form domain, $\dom{\qform}$
is complete under the form norm $\|\cdot\|_{\qform}:=(\qform[\cdot,\cdot]+\|\cdot\|_{L^{2}(\stardom)})^{\frac{1}{2}}$.
Indeed, the latter coincides with the standard Sobolev space norm
on $W^{1,2}(\stardom)$, i.e. $\|\varphi\|_{W^{1,2}(\stardom)}=\|\varphi\|_{\qform}$.
Therefore, the completeness of $W^{1,2}(\stardom)$ implies the completeness
of the form domain.

To prove part \eqref{enu:spectral-properties-of-star-domain-2} we
start by noting that according to Lemma \ref{lem:classC} the boundary
$\partial\stardom$ is of class $C$. This is equivalent to $\stardom$
having the segment property \cite[Theorem~V.4.4]{Edmunds:1987} (see
also \cite[Definition~2.1]{Agmon_book65}). The segment property of
$\stardom$ implies that the Neumann Laplacian, $\lapstar$ has a
compact resolvent \cite[Corollary~1 of Theorem~XIII.75]{ReedSimon_volume4}
and this is equivalent to the discreteness of the spectrum \cite[Theorem~XIII.64]{ReedSimon_volume4}.
\end{proof}
\begin{rem}
Note that the operator $\lapstar$ acts as the standard (negative)
weak Laplacian, as can be shown by using $C_{0}^{\infty}(\stardom)$
functions to cut away the cusp and integrating by parts.
\end{rem}

\begin{rem}
The statements of Proposition \ref{prop:spectral-properties-of-star-domain}
analogously hold for the domain $\lensdom$ as well. Indeed, the boundary
$\partial\lensdom$ possesses no cusp and standard Lipschitz domain
arguments can be applied to show such statements.
\end{rem}

We now investigate the operator domain $\dom{\lapstar}$ in more detail
and show that the restrictions $\left.f_{a,b}\right|_{\stardom}$
belong to $\dom{\lapstar}$, which justifies the definition of the
spectral position for $\stardom$.
\begin{prop}
\label{prop:domain_of_star_Laplacian} ~
\begin{enumerate}
\item \label{enu:domain_of_star_Laplacian-1} The operator domain $\dom{\lapstar}$
satisfies 
\begin{equation}
\dom{\lapstar}\subset\left\{ f\in W^{2,2}(\stardom):\ \left.\partial_{\nu}f\right|_{\partial\stardom}\equiv0\right\} ,\label{domainstar}
\end{equation}
where $\left.\partial_{\nu}f\right|_{\partial\stardom}$ is the normal
derivative.
\item \label{enu:domain_of_star_Laplacian-2} Every separable eigenfunction
$f_{a,b}$ satisfies $\left.f_{a,b}\right|_{\stardom}\in\dom{\lapstar}$.\\
Hence, $\left.f_{a,b}\right|_{\stardom}$ is an eigenfunction of $\lapstar$. 
\end{enumerate}
\end{prop}

\begin{proof}
Part \eqref{enu:domain_of_star_Laplacian-1}. We start by showing
$\dom{\lapstar}\subset W^{2,2}(\stardom)$. Let $f\in\dom{\lapstar}$
and denote $g:=\lapstar f$. We may use \cite[Proposition~8.3.2]{MazPob_book97}
to conclude that there is a unique $W^{1,2}(\stardom)$ solution $\psi$
(up to an additive constant function) for the equation $g=\lapstar(\psi)$.
Explicitly, for the application of \cite[Proposition~8.3.2]{MazPob_book97}
we take $q=q'=1$, $l=1$, $a_{\alpha,\beta}(x)\equiv\delta_{\alpha,\beta}$
(Kronecker delta function) and verify that the assumption on the boundary
near the outer cusp (peak) is satisfied since $\gamma_{a,b}(a)=0$
and $\lim_{x\rightarrow a}\gamma_{a,b}'(x)=0$ (see (\ref{eq:boundary_of_star_domain})).
Next, we use an elliptic regularity result of \cite[Remark~3.3.3]{Grisvard_book11}
to conclude that the equation $g=\lapstar(\psi)$ has a unique $W^{2,2}(\stardom)$
solution (up to an additive constant function). From $W^{2,2}(\stardom)\subset W^{1,2}(\stardom)$
and the uniqueness of the $W^{1,2}(\stardom)$ solution we conclude
that those solutions are the same and since $f\in\dom{\lapstar}\subset W^{1,2}(\stardom)$
we get that this unique solution is $\psi=f$ and $f\in W^{2,2}(\stardom)$.
To apply the regularity result of \cite[Remark~3.3.3]{Grisvard_book11}
we need to verify that the condition imposed there on the boundary
is satisfied. In terms of the notations of \cite[Remark~3.3.3]{Grisvard_book11}
we have $-\phi_{1}(x)=\phi_{2}(x)=\gamma_{a,b}(a-x)$ and the condition
on these functions may be verified with the aid of (\ref{asymp}).

All is left to show is that $\left.\partial_{\nu}f\right|_{\partial\stardom}\equiv0$.
This is done by employing standard localization techniques, as follows.
Let $z\in\partial\stardom$ not identical to a cusp point (i.e., $z\notin\left\{ (0,a),(0,-a)\right\} $).
Then there exists a neighborhood of $z$, say a disc $B\subset\rz^{2}$
which does not contain any of the cusp points of $\stardom$. Choose
a localizing non-negative $C^{\infty}$ function such that $\supp\thinspace\tilde{\eta}=\overline{B}$.
Denote $\eta:=\left.\tilde{\eta}\right|_{\stardom}$.

For all $f\in\dom{\lapstar}$ and $\varphi\in W^{1,2}(\stardom)$
we have 
\begin{equation}
\begin{aligned}\langle\lapstar f,\eta\varphi\rangle_{L^{2}(\stardom)} & =q[f,\eta\varphi]\\
 & =\int_{\stardom}\langle\nabla f,\nabla(\eta\varphi)\rangle_{\C^{2}}\ud x\\
 & =\int_{\supp\thinspace\eta}\langle\nabla f,\nabla(\eta\varphi)\rangle_{\C^{2}}\ud x\\
 & =-\int_{\supp\thinspace\eta}\overline{\mathrm{div}\nabla f}\eta\varphi\ud x+\int_{\partial\left(B\cap\stardom\right)}(\partial_{\nu}f)(\eta\varphi)\ud\sigma\\
 & =-\int_{\stardom}\overline{\mathrm{div}\nabla f}\eta\varphi\ud x+\int_{\partial\left(B\cap\stardom\right)}(\partial_{\nu}f)(\eta\varphi)\ud\sigma\\
 & =\langle\lapstar f,\eta\varphi\rangle_{L^{2}(\stardom)}+\int_{\partial\left(B\cap\stardom\right)}(\partial_{\nu}f)(\eta\varphi)\ud\sigma,
\end{aligned}
\label{eq:Neumann_bc_proof}
\end{equation}
where we stick to the $\mathrm{div}\nabla$ notation and do not use
the standard Laplacian notation in order to distinguish this from
$\lapstar$.

From (\ref{eq:Neumann_bc_proof}) we get that the boundary integral
vanishes. By calculus of variations we conclude that $\partial_{\nu}f|_{\partial\left(B\cap\stardom\right)}=0$
since $\{\eta\varphi|_{\partial\left(B\cap\stardom\right)}:~\varphi\in W^{1,2}(B\cap\stardom)\}$
is dense in $L^{2}\left(\partial\left(B\cap\stardom\right)\right)$.
In particular $\partial_{\nu}f(z)=0$, as required.

To prove part (\ref{enu:domain_of_star_Laplacian-2}) we first truncate
the domain $\stardom$ to remove the cusps. To this end we define
the following family of auxiliary domain
\[
\Omega_{\delta}:=\left\{ (x,y)\in\stardom:\ |x|<a-\delta\right\} ,
\]
and notice that for every $\delta>0$, $\Omega_{\delta}$ is a Lipschitz
domain. Let $\psi\in C^{\infty}(\overline{\stardom})\subset W^{1,2}(\stardom)$
be an arbitrary test function. Then

\begin{align}
q[f_{a,b},\psi] & =\lim_{\delta\rightarrow0}\int_{\Omega_{\delta}}\langle\nabla f_{a,b},\nabla\psi\rangle_{\C^{2}}\ud x\label{eq:form_and_Lap_eigenfunction}\\
 & =\lim_{\delta\rightarrow0}\left[-\int_{\Omega_{\delta}}(\mathrm{div}\nabla f_{a,b})\psi\ud x+\int_{\partial\Omega_{\delta}\setminus\partial\Omega}(\partial_{\nu}f_{a,b})\psi\ud\sigma\right]\nonumber \\
 & =-\int_{\stardom}(\mathrm{div}\nabla f_{a,b})\psi\ud x,\nonumber 
\end{align}
where moving to the last line we have used that both $f_{a,b}$ and
$\mathrm{div}\nabla f_{a,b}$ are bounded, so that the first integral
in the limit converges to an integral over $\stardom$ and the boundary
integral vanishes in the limit.

As is mentioned in the proof of Proposition \ref{prop:spectral-properties-of-star-domain},(\ref{enu:spectral-properties-of-star-domain-2})
the domain $\stardom$ has the segment property. This implies that
$C^{\infty}(\overline{\stardom})$ is dense in $W^{1,2}(\stardom)$
\cite[Theorem 2.1]{Agmon_book65}. Hence we get from (\ref{eq:form_and_Lap_eigenfunction})
that
\begin{equation}
\forall\psi\in W^{1,2}(\stardom)~,\quad q[f_{a,b},\psi]=-\int_{\stardom}(\mathrm{div}\nabla f_{a,b})\psi\ud x.\label{eq:form_and_Lap_eigenfunction_conclusion}
\end{equation}
Since $\dom{\qform}=W^{1,2}(\stardom)$ we get from (\ref{eq:form_and_Lap_eigenfunction_conclusion})
that $f_{a,b}|_{\stardom}\in\dom{\lapstar}$ and $\lapstar f_{a,b}=-\mathrm{div}\nabla f_{a,b}=\lambda_{a,b}f_{a,b}$.
\end{proof}
\begin{rem}
\label{rem:Lens_domain_spectral_prop}The Neumann Laplacian on the
lens domain, $\lensdom$ satisfies similar properties, as can be shown
with analogous arguments. In fact, using the Lipschitz property of
the boundary of $\lensdom$, one can employ standard arguments to
show equality in \eqref{domainstar}, explicitly characterizing the
operator domain.
\end{rem}

\section{Symmetry based analysis towards proof of Theorem \ref{thm:Spectral-pos-separable},(\ref{enu:thm-spectral-pos-star-is-one})
\label{sec:Symmetry}}

In the previous section we have shown that there exists a natural
self-adjoint Neumann Laplacian on $\stardom$, which we denote by
$\lapstar$ and that the spectrum of $\lapstar$ is purely discrete
(Proposition \ref{prop:spectral-properties-of-star-domain}). Next,
we describe a spectral decomposition of $\lapstar$ based on the symmetry
of $\stardom$, which would eventually lead to the proof of Theorem
\ref{thm:Spectral-pos-separable}, (\ref{enu:thm-spectral-pos-star-is-one}).

\begin{figure}[h]
\centering{}\includegraphics[scale=0.7]{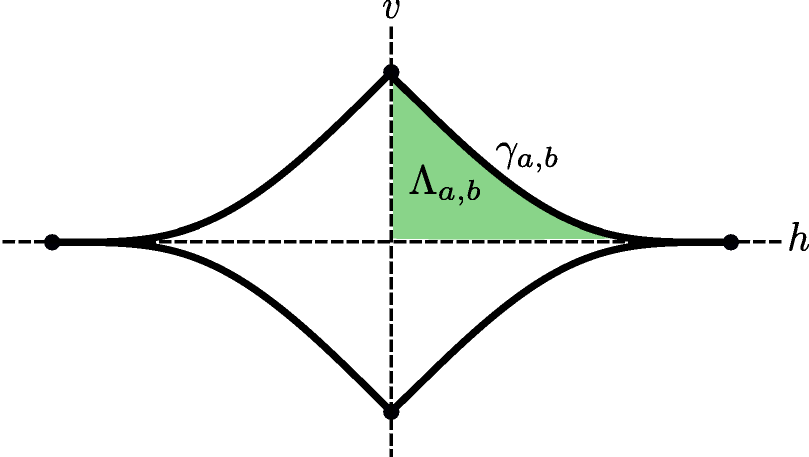} \caption{The star-like domain, $\protect\stardom$ with its two symmetry axes,
$h,~v$. The upper-right quarter, $\protect\quarterdom$, is shaded.}
\label{Fig:quarter_of_star} 
\end{figure}

The domain $\stardom$ has two symmetry axes, horizontal and vertical,
\begin{equation}
h:=\left\{ (x,y):~~y=0\right\} \quad\quad\textrm{and}\quad\quad v:=\left\{ (x,y):~~x=0\right\} \label{eq:horizontal_and_vertical_axes}
\end{equation}
and those dissect $\stardom$ into four quarters (see Figure \ref{Fig:quarter_of_star}).

\begin{figure}[h]
\centering{}\includegraphics[scale=0.7]{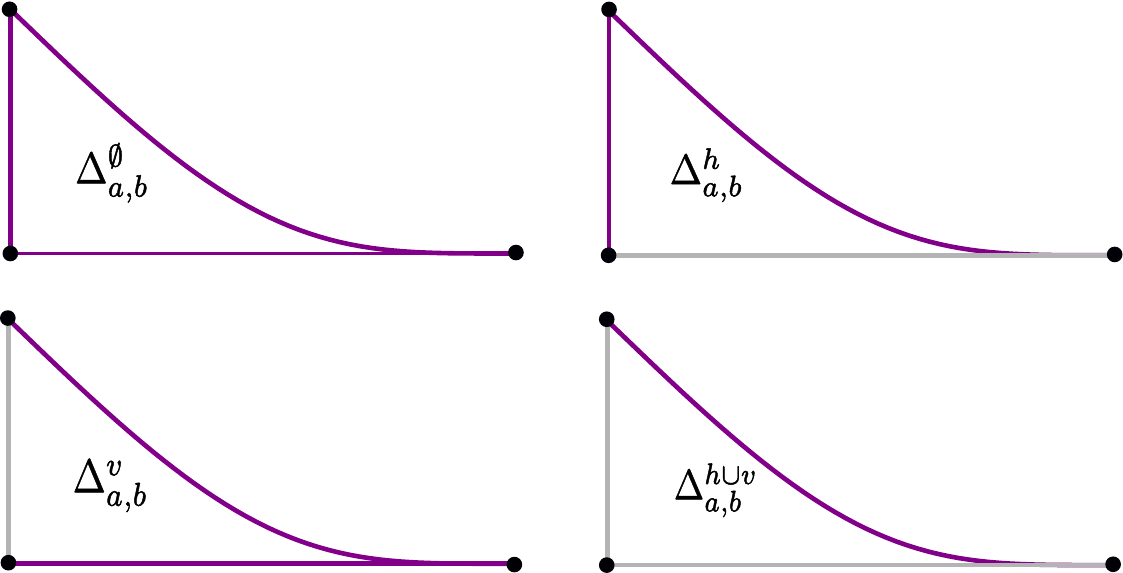} \caption{The four Laplacians $\protect\lap_{a,b}^{\emptyset}$, $\protect\laph$
, $\protect\lapv$ and $\protect\lap_{a,b}^{h\cup v}$, all defined
on $\protect\quarterdom$ and differ only in boundary conditions.
Purple color indicates Neumann conditions and gray stands for Dirichlet.}
\label{Fig:quotients} 
\end{figure}

We denote by $\quarterdom$ the upper-right quarter,
\begin{equation}
\quarterdom=\left\{ (x,y):\ 0<x<a,~0<y<\gamma_{a,b}(x)\right\} ,\label{eq:quarter_star_domain}
\end{equation}
and note that $\quarterdom$ is bounded by $\gamma_{a,b}$, $h$ and
$v$, i.e., 
\begin{equation}
\partial\quarterdom\subset h~\cup~v~\cup~\gamma_{a,b}.\label{eq:quarter_star_boundary}
\end{equation}
Next, we introduce four Laplace-Beltrami operators on $\quarterdom$,
which differ only in their boundary conditions. We denote those Laplacians
by $\lap_{a,b}^{\emptyset}$, $\laph$ , $\lapv$ and $\lap_{a,b}^{h\cup v}$,
and also use the notation $\lap_{a,b}^{D}$ when we want to refer
to either of them ($D\in\left\{ \emptyset,~h,~v,~h\cup v\right\} $).
The superscripts of those Laplacians indicate which part of the boundary
$\partial\quarterdom$ serves as the Dirichlet boundary, whereas the
rest of the boundary is taken to be Neumann (see Figure \ref{Fig:quotients}).

Using Proposition 2 in \cite{ParBan_jga10} or Property II,(3) in
\cite{BanBerJoyWen_arXiv} we obtain the following spectral decomposition\footnote{To apply the theory in \cite{ParBan_jga10,BanBerJoyWen_arXiv} for
our case, we take the group to be $C_{2}\times C_{2}$ (the direct
product of two copies of the cyclic group, $C_{2}$) with its regular
representation.}
\begin{equation}
\spec{\lapstar}=\spec{\lap_{a,b}^{\emptyset}}\cup\spec{\lap_{a,b}^{h}}\cup\spec{\lap_{a,b}^{v}}\cup\spec{\lap_{a,b}^{h\cup v}}.\label{eq:Spectral_decomposition_of_star}
\end{equation}
The equality above holds also when taking into account the multiplicities
of eigenvalues on both sides.

The spectral decomposition (\ref{eq:Spectral_decomposition_of_star})
may be also understood on an intuitive level, as follows. Since the
domain $\stardom$ is symmetric with respect to reflection along both
$h$ and $v$ we get that that the Neumann Laplacian, $\lapstar$
commutes with each of those symmetries. As a result, $\lapstar$ posseses
a complete set of eigenfunctions which respects this symmetry. Namely,
each eigenfunction in this basis is either symmetric or anti-symmetric
with respect to $h$ and either symmetric or anti-symmetric with respect
to $v$. Each eigenfunction therefore belongs to one of four symmetry
classes and its corresponding eigenvalue belongs to either of the
spectra $\spec{\lap_{a,b}^{\emptyset}},~\spec{\lap_{a,b}^{h}},~\spec{\lap_{a,b}^{v}},~\spec{\lap_{a,b}^{h\cup v}}$.

Now, Theorem \ref{thm:Spectral-pos-separable},(\ref{enu:thm-spectral-pos-star-is-one})
follows when combining the spectral decomposition (\ref{eq:Spectral_decomposition_of_star})
together with the following two propositions (the propositions are
proven in the next section).
\begin{prop}
\label{prop:Two-symmetry-classes} 
\[
\lambda_{1}(\lapstar)\notin\spec{\lap_{a,b}^{\emptyset}}\cup\spec{\lap_{a,b}^{h\cup v}}
\]
\end{prop}

\begin{prop}
\label{prop:eigenvalue-inequality-vert-horiz} There exists $c>1$
such that if $\nicefrac{a}{b}>c$ then
\begin{equation}
\lambda_{1}(\lap_{a,b}^{v})<\lambda_{1}(\lap_{a,b}^{h}).\label{eq:eigenvalue-inequality-vert-horiz}
\end{equation}

\vspace{4mm}
\end{prop}

\begin{proof}
[Proof of Theorem \ref{thm:Spectral-pos-separable},(\ref{enu:thm-spectral-pos-star-is-one})]

From the spectral decomposition (\ref{eq:Spectral_decomposition_of_star})
and Proposition \ref{prop:Two-symmetry-classes} we get that either
$\lambda_{1}(\lapstar)=\lambda_{1}(\lap_{a,b}^{h})$ or $\lambda_{1}(\lapstar)=\lambda_{1}(\lap_{a,b}^{v})$.
Then, by (\ref{eq:eigenvalue-inequality-vert-horiz}) we deduce that
actually $\lambda_{1}(\lapstar)=\lambda_{1}(\lap_{a,b}^{v})$ for
$\nicefrac{a}{b}>c$.

Now, assume $\nicefrac{a}{b}>c$ and consider the eigenfunction $f_{a,b}$
corresponding to $\lambda_{a,b}$ (see (\ref{eq:Torus-eigenfunction})).
The restriction $\left.f_{a,b}\right|_{\stardom}$ is symmetric with
respect to $h$ and anti-symetric with respect to $v$ and hence $\lambda_{a,b}\in\spec{\lap_{a,b}^{v}}$.
Furthermore, the restriction to the quarter star, $\left.f_{a,b}\right|_{\quarterdom}$
has a single nodal domain so it is the first eigenfunction of $\lapv$
(this follows from Courant's bound \cite{Courant23} together with
orthogonality of eigenfunctions), i.e., $\lambda_{a,b}=\lambda_{1}(\lapv)$.
Combining this with $\lambda_{1}(\lapstar)=\lambda_{1}(\lap_{a,b}^{v})$
which we obtained above, we get $\lambda_{a,b}=\lambda_{1}(\lapstar)$,
so that $N_{\stardom}(\lambda_{a,b})=1$, as required.

Finally, the simplicity of $\lambda_{a,b}$ as an eigenvalue of $\lapstar$
also follows from the arguments above. We got that $\lambda_{1}(\lapstar)=\lambda_{1}(\lap_{a,b}^{v})$
and also that $\lambda_{1}(\lapstar)\notin\spec{\lap_{a,b}^{\emptyset}}\cup\spec{\lap_{a,b}^{h}}\cup\spec{\lap_{a,b}^{h\cup v}}$.
By the spectral decomposition (\ref{eq:Spectral_decomposition_of_star}),
this means that $\lambda_{1}(\lapstar)$ may be a multiple eigenvalue
only if $\lambda_{1}(\lap_{a,b}^{v})$ itself is a multiple eigenvalue
of $\lap_{a,b}^{v}$. But $\lambda_{1}(\lap_{a,b}^{v})$ is the lowest
eigenvalue of $\lap_{a,b}^{v}$ and hence must be simple.
\end{proof}

\section{Proofs of Propositions \ref{prop:Two-symmetry-classes} and \ref{prop:eigenvalue-inequality-vert-horiz}
\label{sec:Rearrangement}}
\begin{proof}
[Proof of Proposition \ref{prop:Two-symmetry-classes}]
\begin{figure}[h]
\centering{}\includegraphics[width=0.8\textwidth]{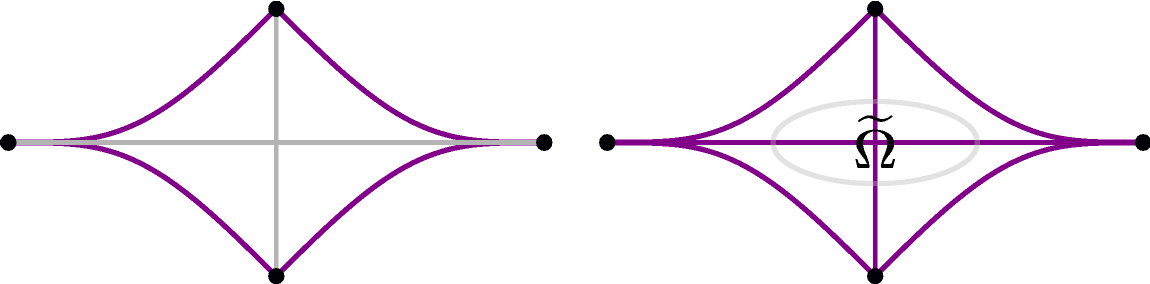}
\caption{Depicting properties of particular eigenfunctions of $\protect\lapstar$.
Grey curves indicate the nodal set of the eigenfunciton and purplue
curves mark vanishing of the normal derivative.\protect \\
\uline{Left}: An eigenfunction which is anti-symmetric both with
respect to $h$ and $v$ (so that the corresponding eigenvalue satisfies
$\lambda\in\protect\spec{\protect\lap_{a,b}^{h\cup v}}$).\protect \\
\uline{Right}: An eigenfunction which is symmetric both with respect
to $h$ and $v$ (so that the corresponding eigenvalue satisfies $\lambda\in\protect\spec{\protect\lap_{a,b}^{\emptyset}}$).}
\label{Fig:two_symmetry_types} 
\end{figure}
First, assume by contradiction that $\lambda_{1}(\lapstar)\in\spec{\lap_{a,b}^{h\cup v}}$.
This means that there exists an eigenfunction $f$ corresponding to
$\lambda_{1}(\lapstar)$, such that $f$ is anti-symmetric both with
respect to the $h$ axis and the $v$ axis. In particular, the nodal
set of $f$ contains both $h$ and $v$ and therefore $f$ has at
least four nodal domains (see Figure \ref{Fig:two_symmetry_types}(Left)).
This brings to a contradiction, since by Courant's bound \cite{Courant23},
each eigenfunction which corresponds to $\lambda_{1}(\lapstar)$ has
at most two nodal domains (note that $\lambda_{0}(\lapstar)=0$, so
that $\lambda_{1}(\lapstar)$ is the second lowest eigenvalue).

Next, assume by contradiction that $\lambda_{1}(\lapstar)\in\spec{\lap_{a,b}^{\emptyset}}$.
This means that there exists an eigenfunction $f$ corresponding to
$\lambda_{1}(\lapstar)$, such that $f$ is symmetric both with respect
to the $h$ axis and the $v$ axis. As above, $f$ must have exactly
two nodal domains (it cannot have a single nodal domain, as $f$ must
be orthogonal to the constant eigenfunction). The only possibility
for $f$ to be symmetric as above and contain two nodal domains is
if $f$ has a single closed nodal line which is symmetric with respect
both to the $h$ axis and the $v$ axis (see Figure \ref{Fig:two_symmetry_types}(Right)).
This brings to a contradiction, as the first non-trivial Neumann eigenfunction
cannot have a closed nodal line \cite{Pleijel_cpam56} (see also \cite{Payne_siam67}).
There is a simple argument for that, which we bring here for completeness.
We denote the interior nodal domain of $f$ (the one which does not
touch the boundary) by $\widetilde{\Omega}$. In particular we get
that $\left.f\right|_{\widetilde{\Omega}}$ is a Dirichlet eigenfunction
of $\widetilde{\Omega}$ whose eigenvalue is $\lambda_{1}(\lapstar)$.
Furthermore, $\left.f\right|_{\widetilde{\Omega}}$ has a single nodal
domain and hence it is the first Dirichlet eigenfunction of $\widetilde{\Omega}$,
i.e., $\lambda_{1}(\lapstar)=\lambda_{1}^{(D)}(\widetilde{\Omega})$
(this follows from Courant's bound \cite{Courant23} together with
eigenfunction orthogonality). If we now denote by $\lambda_{1}^{(D)}(\stardom)$
the first Dirichlet eigenvalue of $\stardom$, we get 
\begin{equation}
\lambda_{1}^{(D)}(\stardom)\leq\lambda_{1}^{(D)}(\widetilde{\Omega})=\lambda_{1}(\lapstar)<\lambda_{1}^{(D)}(\stardom),\label{eq:contradicting_inequalities}
\end{equation}
where the left inequality above follows from the monotonicity of Dirichlet
eigenvalues and the right inequality appears already in \cite{Polya_jmp52}
(see also \cite{Payne_siam67} and note that it is actually part of
a more general interlacing property of Dirichlet and Neumann eigenvalues
\cite{Filonov_aia04,Friedlander_rma91}). Overall, (\ref{eq:contradicting_inequalities})
is a contradiction, which means that $f$ cannot be symmetric with
respect to both the $v$ and the $h$ axes and $\lambda_{1}(\lapstar)\notin\spec{\lap_{a,b}^{\emptyset}}.$

\end{proof}
Next, to prove Proposition \ref{prop:eigenvalue-inequality-vert-horiz}
we need to compare the first eigenvalues of the operators $\lapv$
and $\laph$. We do so by using a sector as an auxiliary domain and
proving that its first eigenvalue lies in between $\lambda_{1}(\lapv)$
and $\lambda_{1}(\laph)$. This is done in Lemmata \ref{lem:eigenvalue-inequality-sect-horiz}
and \ref{lem:eigenvalue-inequality-vert-sect} and for this purpose
we denote

\begin{equation}
\sector:=\left\{ \left(r\cos(\phi),~r\sin(\phi)\right):\quad0<r<R,\ \left|\phi\right|<\nicefrac{\pi}{8}\right\} .
\end{equation}
In particular, we choose $R$ to be such that the sector area equals
the area of $\stardom,$i.e., $\left|S_{R}\right|=\frac{\pi}{8}R^{2}=\left|\quarterdom\right|$.
We consider the Laplacian on $\sector$ with
\[
\textrm{Neumann boundary conditions on \ensuremath{\left\{  \left(r\cos(\phi),~r\sin(\phi)\right):~0<r<R,\ |\phi|=\nicefrac{\pi}{8}\right\} } }
\]
\[
\textrm{and Dirichlet boundary conditions on }\left\{ \left(R\cos(\phi),~R\sin(\phi)\right):~|\phi|<\nicefrac{\pi}{8}\right\} .\quad\quad\quad\quad
\]
We denote the first eigenvalue of this Laplacian on the sector by
$\lambda_{1}(\sector)$.
\begin{lem}
\label{lem:eigenvalue-inequality-sect-horiz} We have
\begin{equation}
\lambda_{1}(\sector)<\lambda_{1}(\laph).\label{eq:eigenvalue-inequality-sect-horiz}
\end{equation}
\end{lem}

\begin{proof}
We employ a spectral isoperimetric inequality \cite[SATZ 3]{Bandle_cmh71}
(see also \cite[Theorem 3.9]{Bandle_book80}) to get 
\begin{equation}
\lambda_{1}(\laph)\geq\frac{\nicefrac{\pi}{4}}{2\left|\quarterdom\right|}j_{0}^{2},\label{eq:eigenvalue-inequality-sect-horiz-1}
\end{equation}
where $j_{0}\approx2.4048$ is the first zero of $J_{0}$, the zeroth
Bessel function. The value $\nicefrac{\pi}{4}$ in the RHS of (\ref{eq:eigenvalue-inequality-sect-horiz-1})
is determined in \cite[Theorem 3.9]{Bandle_book80} as the so called
'rotation' of the Neumann boundary of $\laph$ (which is the curve
$v\cup\gamma_{a,b}$). The definition of the rotation of a curve is
given in \cite[Section 2.3]{Bandle_book80} and in our case, due to
the concavity of $\gamma_{a,b}$ it simply equals to the opening angle
of $\partial\quarterdom$ at $(0,b)$. We note that $\lambda_{1}(\sector)$
is equal to the first Dirichlet eigenvalue of the disc of radius $R$.
Namely, 
\begin{equation}
\lambda_{1}(\sector)=\left(\frac{j_{0}}{R}\right)^{2}=j_{0}^{2}\frac{\nicefrac{\pi}{8}}{\text{\ensuremath{\left|\quarterdom\right|}}},\label{eq:First_sector_eigenvalue}
\end{equation}
which is the RHS of (\ref{eq:eigenvalue-inequality-sect-horiz-1}),
so we get $\lambda_{1}(\sector)\leq\lambda_{1}(\laph)$. This inequality
is actually strict since $\quarterdom$ is not a circular sector \cite[SATZ 3]{Bandle_cmh71}.
\end{proof}
\begin{lem}
\label{lem:eigenvalue-inequality-vert-sect} There exists $c>1$ such
that if $\frac{a}{b}>c$ then 
\begin{equation}
\lambda_{1}(\lapv)<\lambda_{1}(\sector).\label{eq:eigenvalue-inequality-vert-sect}
\end{equation}
\end{lem}

\begin{proof}
We start by observing that $\lambda_{1}(\lapv)=\frac{\pi^{2}}{4}\left(\frac{1}{a^{2}}+\frac{1}{b^{2}}\right)$
(see (\ref{eq:Torus-eigenvalue})). Indeed, $\frac{\pi^{2}}{4}\left(\frac{1}{a^{2}}+\frac{1}{b^{2}}\right)$
is the eigenvalue corresponding to the eigenfunction $f_{a,b}(x,y)=\sin\left(\frac{\pi}{2a}x\right)\cos\left(\frac{\pi}{2b}y\right)$
on $\stardom$ (see (\ref{eq:Torus-eigenfunction})). The restriction
$\left.f_{a,b,}\right|_{\quarterdom}$ is in the domain of the operator
$\lapv$ since it fulfills Dirichlet boundary conditions at $v$ and
Neumann boundary conditions at $h$ (and arguing similarly to Proposition
\ref{prop:domain_of_star_Laplacian},(\ref{enu:domain_of_star_Laplacian-2}))).
Furthermore, $\left.f_{a,b,}\right|_{\quarterdom}$ has a single nodal
domain and hence it is the first eigenfunction, so that $\lambda_{1}(\lapv)=\frac{\pi^{2}}{4}\left(\frac{1}{a^{2}}+\frac{1}{b^{2}}\right)$.

Noting $\lambda_{1}(\sector)=j_{0}^{2}\frac{\nicefrac{\pi}{8}}{\text{\ensuremath{\left|\quarterdom\right|}}}$
(see (\ref{eq:First_sector_eigenvalue})), means that (\ref{eq:eigenvalue-inequality-vert-sect})
is equivalent to 
\begin{equation}
\left(\frac{1}{a^{2}}+\frac{1}{b^{2}}\right)\left|\quarterdom\right|<\frac{(j_{0})^{2}}{2\pi}.\label{eq:bound_on_star_area}
\end{equation}
We defer this last part of the proof to Lemma \ref{lem:bound_on_star_area},
where it is shown that there exists $c>1$ such that (\ref{eq:bound_on_star_area})
holds whenever $\frac{a}{b}>c$.
\end{proof}
Proposition \ref{prop:eigenvalue-inequality-vert-horiz} now follows
as an immediate implication of Lemmata \ref{lem:eigenvalue-inequality-sect-horiz}
and \ref{lem:eigenvalue-inequality-vert-sect}.

\section{The area-to-perimeter ratio and numerics \label{sec:Numerics}}

In this section we introduce a geometric parameter (a normalized area-to-perimeter
ratio) which promotes a further investigation of the spectral position
from a numerical perspective.

\subsection{The normalized area-to-perimeter ratio}
\begin{defn}
\label{def:Area-to-Perimeter-Manifolds} \cite{EGJS07} Let $f$ be
a Morse eigenfunction corresponding to the eigenvalue $\lambda$ and
let $\Omega$ be a Neumann domain of $f$. We define the normalized
area to perimeter ratio of $\Omega$ by
\[
\rho(\Omega):=\frac{\left|\Omega\right|}{\left|\partial\Omega\right|}\sqrt{\lambda},
\]
with $\left|\Omega\right|$ being the area of $\Omega$ and $\left|\partial\Omega\right|$
the total length of its perimeter.\\

This parameter was originally introduced in \cite{EGJS07} for the
study of nodal domain geometry. If we consider merely the area to
perimeter ratio (without normalizing by the eigenvalue), then the
value $\frac{\left|\Omega\right|}{\left|\partial\Omega\right|}$ has
an interesting geometric meaning \cite{MazdeMZoi_jmp14}, being equal
to the mean chord length of the two-dimensional shape $\Omega$ (up
to a multiplicative factor of $\frac{1}{\pi}$). The mean chord length
is defined as follows; consider all the parallel chords in a chosen
direction and take their average length; the uniform average of this
value over all directions is the mean chord length\footnote{We thank John Hannay for pointing out this interesting geometrical
meaning to us.}.

Interestingly, the value of $\rho(\Omega)$ is also connected to the
spectral position of the Neumann domain, $\Omega$.
\end{defn}

\begin{prop}
\label{prop:Rho-upper-bound}Let $f$ be a Morse eigenfunction corresponding
to eigenvalue $\lambda$. Let $\Omega$ be a Neumann domain of $f$.
We have 
\begin{enumerate}
\item $\rho(\Omega)\leq\sqrt{2N_{\Omega}(\lambda)}$.\label{enu:prop-Rho-upper-bound-manifold-1}
\item if $N_{\Omega}(\lambda)=1$ then $\rho(\Omega)\leq\frac{j'}{2}\approx0.9206$\label{enu:prop-Rho-upper-bound-manifold-2}
\item if $N_{\Omega}(\lambda)=2$ then $\rho(\Omega)\leq\frac{j'}{\sqrt{2}}\approx1.3019$,
\label{enu:prop-Rho-upper-bound-manifold-3}
\end{enumerate}
where $j'\approx1.8412$ is the first zero of the derivative of $J_{1},$
the first Bessel function.

\end{prop}

\begin{proof}
We write $\rho(\Omega)=\frac{\left|\Omega\right|}{\left|\partial\Omega\right|}\sqrt{\lambda}=\frac{\sqrt{\left|\Omega\right|}}{\left|\partial\Omega\right|}\sqrt{\lambda\left|\Omega\right|}$.
The first factor in this product, is bounded from above by the classical
geometric isoperimetric inequality $\frac{\sqrt{\left|\Omega\right|}}{\left|\partial\Omega\right|}\leq\frac{1}{2\sqrt{\pi}}$
\cite{Federer_book69},\cite[Theorem 14.1]{Maggi_book12}. An equality
occurs if and only is $\Omega$ is a disc. The second factor is bounded
from above in terms of the spectral position \cite{Kro_jfa92}
\begin{equation}
\lambda\left|\Omega\right|\leq8\pi N_{\Omega}(\lambda),\label{eq:Kroger_inequality}
\end{equation}
and combining both we get the first bound of the proposition. In the
particular cases of $N_{\Omega}(\lambda)=1$ or $N_{\Omega}(\lambda)=2$
the bound (\ref{eq:Kroger_inequality}) may be improved as follows.

By the Szegö-Weinberger inequality \cite{Szego_jrma54,Weinberger_jrma56}
we have that if $N_{\Omega}(\lambda)=1$ then 
\begin{equation}
\lambda\left|\Omega\right|\leq\pi\left(j'\right){}^{2}.\label{eq:Szego-Weinberger}
\end{equation}
Combining this bound with the geometric isoperimetric inequality proves
the second part of the proposition. 

By the Giraurd-Nadirashvili-Polterovich inequality \cite{GirNadPol_jdg09}
we have that if $N_{\Omega}(\lambda)=2$ then 
\begin{equation}
\lambda\left|\Omega\right|\leq2\pi\left(j'\right){}^{2}.\label{eq:Giraurd}
\end{equation}
Combining this bound with the geometric isoperimetric inequality proves
the third part of the proposition. 
\end{proof}
The last proposition allows to use numerics to estimate the spectral
position of Neumann domains. We note that the exact value of the spectral
position cannot be easily computed not even numerically. Part of the
difficulty arises since for a general Neumann domain we do not have
an analytic expression of its boundary. So, computing the spectrum
is highly nontrivial. As apposed to that, the area-to-perimeter ratio
is relatively easily computed when the Neumann lines of an eigenfunction
are numerically found. Once calculating $\rho(\Omega)$, the last
proposition allows to deduce that Neumann domains whose $\rho(\Omega)$
value is large enough do not have low spectral positions.

\subsection{Numerical results for separable eigenfunctions}

We use the parameter $\rho$ to further investigate the spectral positions
of Neumann domains of separable eigenfunctions on the torus. As we
have seen in previous sections, a particular separable eigenfunction
has only two congruence classes of Neumann domains, the lens-like
and the star-like (Figure \ref{fig:Torus-with-star-and-lens}). So
each separable eigenfunction have just two possible $\rho$ values,
$\rho(\stardom$) and $\rho(\lensdom)$. These $\rho$ values change
with the eigenfunction and solely depend on the ratio $\nicefrac{a}{b}$,
where $a,b$ are the values which characterize the eigenfunction,
(\ref{eq:Torus-eigenfunction}).

\begin{figure}[h]
\centering{}\includegraphics[width=1\textwidth]{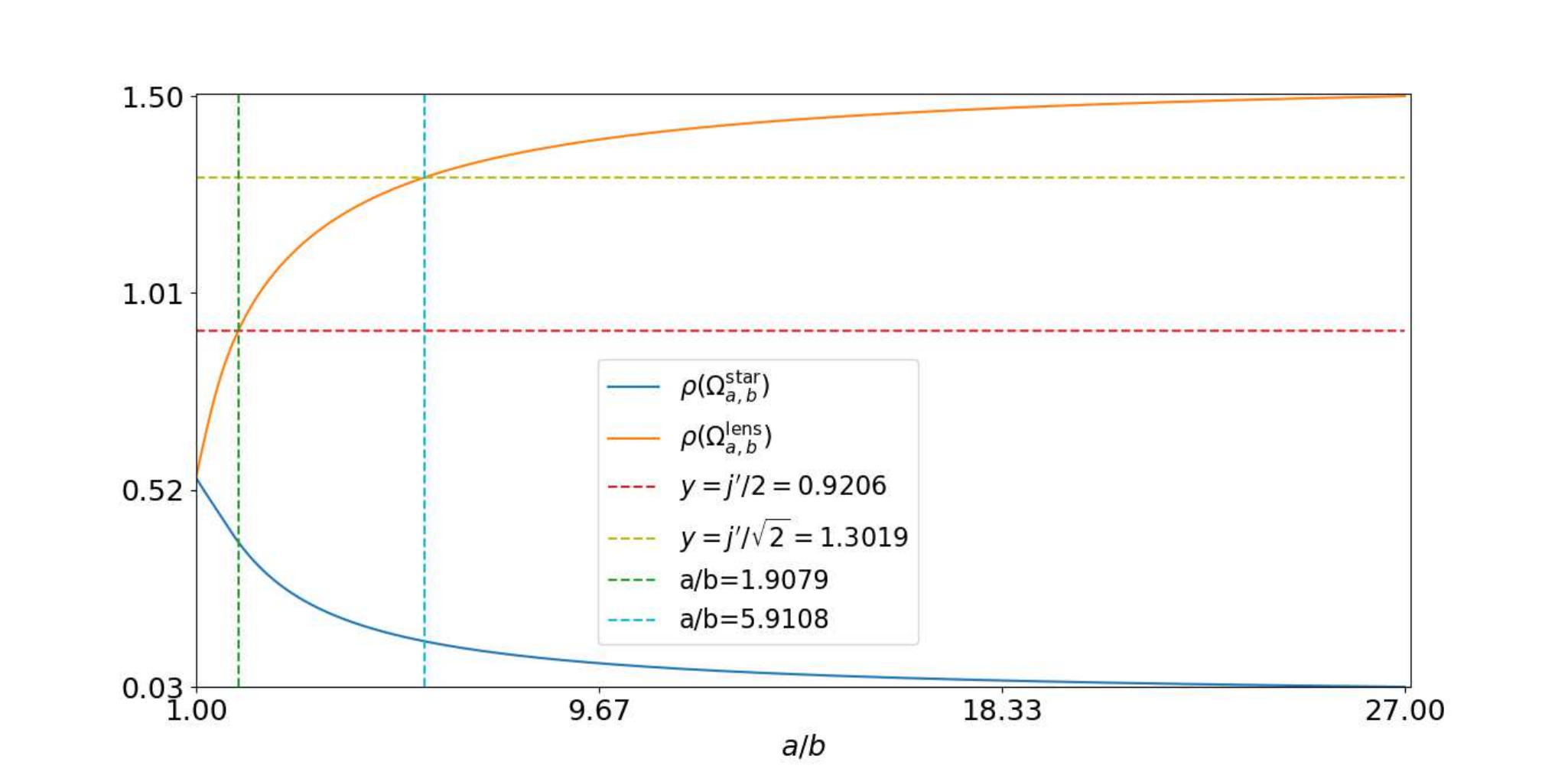}
\caption{The values $\rho(\protect\stardom)$ and $\rho(\protect\lensdom)$
for the two Neumann domains of the separable eigenfunction $f_{a,b}$,
(\ref{eq:Torus-eigenfunction}), as a function of the ratio $\nicefrac{a}{b}$.
The critical $\rho$ values $0.9206,~1.3019$ from Proposition \ref{prop:Rho-upper-bound}
are marked.}
\label{Fig:rho_of_separable} 
\end{figure}
Figure \ref{Fig:rho_of_separable} shows how the values $\rho(\stardom$)
and $\rho(\lensdom)$ depend on $\nicefrac{a}{b}$. In particular,
we observe that $\rho(\lensdom)$ increases with $\nicefrac{a}{b}$.
Using Proposition \ref{prop:Rho-upper-bound} we may conclude that
if $\nicefrac{a}{b}>1.9079$ then $N_{\lensdom}\left(\lambda_{a,b}\right)>1$.
Similarly, if $\nicefrac{a}{b}>5.9108$ then $N_{\lensdom}\left(\lambda_{a,b}\right)>2$.
Note that according to Theorem \ref{thm:Spectral-pos-separable},(\ref{enu:thm-spectral-pos-lens})
$N_{\lensdom}\left(\lambda_{a,b}\right)\rightarrow\infty$ as $\nicefrac{a}{b}\rightarrow\infty$.
The numerical values above allow to slightly refine this result and
estimate the growth rate of $N_{\lensdom}\left(\lambda_{a,b}\right)$
with $\nicefrac{a}{b}$.

\subsection{Numerical results for random waves on the torus\label{subsec:Numerics-random-waves}}

\begin{figure}[H]
\includegraphics[scale=0.65]{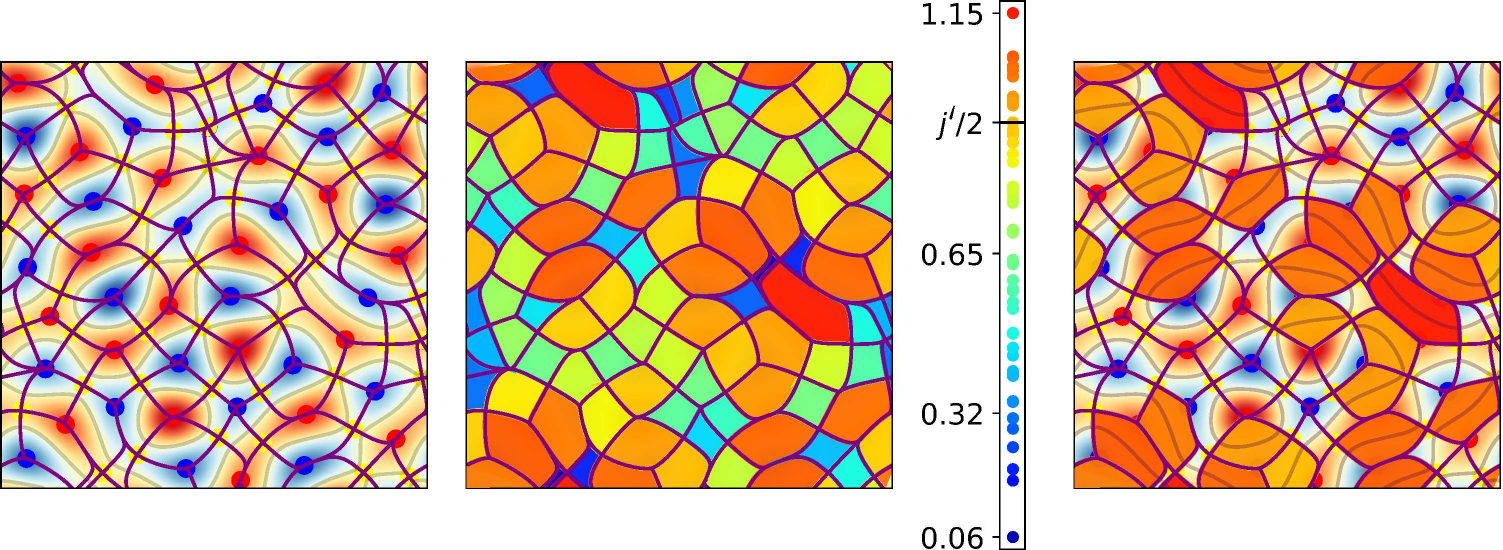}
\caption{\uline{Left}: The Neumann partition of an eigenfunction with $\lambda=25$
on the two-dimensional torus.\protect \\
\uline{Middle}: The Neumann domains colored according to their
$\rho$ value.\protect \\
\uline{Right}: Only Neumann domains for which $\rho(\Omega)>\frac{j'}{2}$
are colored according to their $\rho$ value. The rest are colored
as in the left part of the figure.}
\label{fig:Neumann_domains_and_rho} 
\end{figure}

We start by demonstrating an application of Proposition \ref{prop:Rho-upper-bound}
to estimate spectral positions of arbitrary Neumann domains. As an
example, we show in Figure \ref{fig:Neumann_domains_and_rho} the
Neumann partition of an eigenfunction with $\lambda=25$ on the two-dimensional
torus. On the left part, only the Neumann partition is shown. On the
middle figure each Neumann domain $\Omega$ is colored according to
its $\rho(\Omega)$ value and on the right we keep colored only Neumann
domains with $\rho(\Omega)>\nicefrac{j'}{2}$ (the rest are left colored
blue-red, as in the left figure). From Proposition \ref{prop:Rho-upper-bound},(\ref{enu:prop-Rho-upper-bound-manifold-2})
we deduce that all colored Neumann domains have spectral position
larger than one.

Following this pictorial demonstration, we also calculate the probability
distribution of $\rho(\Omega)$ values. We follow the random wave
model in our computations. We choose a certain non-simple eigenvalue
of the torus. We consider a certain basis of eigenfunctions of this
eigenspace and take linear combinations of those eigenfunctions, where
the coefficients are chosen according to the standard normal distribution.
This describes the random ensemble that we use. We pick approximately
$9000$ eigenfunctions from this ensemble and for each we calculate
$\rho$ values for all of its Neumann domains to obtain the probability
distribution of $\rho(\Omega)$ for this particular eigenvalue. 

\begin{figure}[H]
\includegraphics[width=1\textwidth]{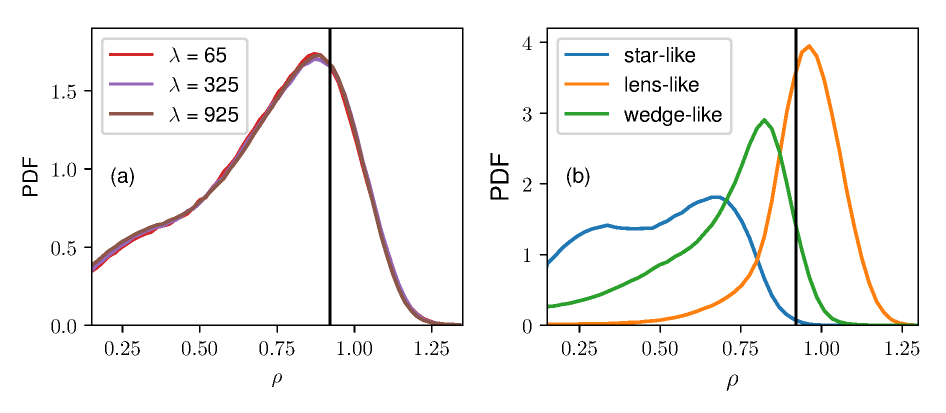}
\caption{\uline{Left}: The probability distribution of $\rho$ values for
Neumann domains of random eigenfunctions drawn from three different
eigenvalues ($\lambda\in\left\{ 65,325,925\right\} $). The vertical
lines marks the bound $\nicefrac{j'}{2}$ from Proposition \ref{prop:Rho-upper-bound},(\ref{enu:prop-Rho-upper-bound-manifold-2}).\protect \\
\uline{Right}: The probability distributions of $\rho$ values
for Neumann domains separated according to the type of the Neumann
domain (star, lens, wedge). The random eigenfunctions are of eigenvalue
$\lambda=925$. \protect \\
The results are drawn according to 2494622 lens-like domains, 2670896
star-like domains and 3283304 wedge-like domains, numerically traced
and analyzed from approximately 9000 individual eigenfunctions.}
\label{fig:rho_prob_dist} 
\end{figure}
The results are shown in Figure \ref{fig:rho_prob_dist}. On the left
we plot the probability distribution of three different eigenvalues
(65, 325, 925) of the flat torus of side length $2\pi$. A first observation
is that a substantial proportion (around $23$ percent) of the Neumann
domains have a $\rho$ value which is larger than $\nicefrac{j'}{2}$
(indicated by a vertical line in the figure), the upper bound in Proposition
\ref{prop:Rho-upper-bound},(\ref{enu:prop-Rho-upper-bound-manifold-2}).
A combination of this numerical observation and the Proposition shows
that at least some $23$ percent of the Neumann domains have spectral
position larger than one. Furthermore, this plots suggests that the
$\rho$ distribution might be independent of the particular eigenvalue.
We currently do not have an analytic explanation to this numeric finding.

Another interesting observation can be made from the right part of
Figure \ref{fig:rho_prob_dist}, which separately shows the $\rho$
distribution of three different types of Neumann domains. Neumann
domains may be classified into three types according to the angles
their boundary forms at the critical points. Neumann lines always
meet perpendicularly at saddle points, whereas at extremal points
Neumann lines might meet either at an angle of $\pi$ or of $0$,
\cite[Proposition 4.1]{McDFul_ptrs13,AloBanBerEgg_Neumann}. If both
angles at minimum and maximum points of a Neumann domain are $0$
we call the Neumann domain star-like. If both angles are $\pi$ we
call it lens-like and if one angle is $0$ and the other is $\pi$
then the Neumann domain is called wedge-like (see right part of Figure
\ref{fig:Neumann-Domains-Torus-Basic}). With this distinction, we
may compute the $\rho$ values for each of the three types above separately
(as before, according to the random wave model). Doing so for random
eigenfunctions of the eigenvalue $925$ results with the right plot
in Figure \ref{fig:rho_prob_dist}. This plot might indicate that
this three-fold classification affects the geometry of Neumann domains
and might suggest a direction for exploring the probability distribution
of $\rho$ and its possible universality. 

We end by pointing out the role numerics play in our work. That the
spectral position may be higher than one is counter-intuitive (see
end of Section \ref{subsec:Spectral-position}) and is stated in Theorem
\ref{thm:Spectral-pos-separable},(\ref{enu:thm-spectral-pos-lens}).
Yet, this result concerns to separable eigenfunctions which are exceptional
in some sense. One may wonder whether it is only for this special
case that the spectral position differs than one or is it more general
than that. The numerics we supply here for random eigenfunctions indicate
that generically there is a non-negligible probability that a Neumann
domain would have a spectral position larger than one. This calls
for further investigations of the Neumann domains spectral positions.

\subsection*{Acknowledgment}

We would like to thank Emanuel Milman for stimulating discussions
and for pointing out helpful references. We thank Gregory Berkolaiko,
Mark Dennis and Michael Levitin for interesting discussions in various
stages of this ongoing work. We thank Luc Hillairet and Graham cox
for pointing out to us the second proof of Theorem \ref{thm:Spectral-pos-separable},(\ref{enu:thm-spectral-pos-lens}).
Band and Egger were supported by ISF (Grant No. 494/14). Taylor was
funded by the Leverhulme Trust Research Programme Grant No. RP2013-K-009.

\appendix

\section{the boundary of $\protect\stardom$ and its area \label{sec:Appendix-area-calculation-of-star}}

We consider a separable eigenfunction $f_{a,b}$ on the torus, (\ref{eq:Torus-eigenfunction}),
and its star-like Neumann domain, $\stardom$. In this appendix we
derive the explicit expression for the boundary of $\stardom$ (Lemma
\ref{lem:gamma_formula}) and show that it is of class C (Lemma \ref{lem:classC}).
This boundary characterization is needed to justify the application
of some Sobolev space analysis (done in Proposition \ref{prop:spectral-properties-of-star-domain}).
Furthermore, we perform here an asymptotic calculation of the $\stardom$
area (Lemma \ref{lem:bound_on_star_area}) which is used in the proofs
of Theorem \ref{thm:Spectral-pos-separable},(\ref{enu:thm-spectral-pos-lens})
(first proof) and Lemma \ref{lem:eigenvalue-inequality-vert-sect}.
\begin{lem}
\label{lem:gamma_formula}We have 
\[
\stardom=\left\{ \left(x,y\right)~:~\left|x\right|<a,~~\left|y\right|<\gamma_{a,b}(x)\right\} ,
\]
where 
\begin{equation}
\gamma_{a,b}(x):=\frac{2b}{\pi}\arcsin\left(\left[\cos\left(\frac{\pi}{2a}x\right)\right]^{\left(\frac{a}{b}\right)^{2}}\right).\label{eq:Lemma_boundary_of_star_domain}
\end{equation}
\end{lem}

\begin{proof}
To prove the Lemma we parameterize the Neumann line which connects
the extremal point, $(a,0)$ to the saddle point, $(0,b)$ (see Figure
\ref{fig:Torus-with-star-and-lens},(ii)) and show that it is given
by (\ref{eq:Lemma_boundary_of_star_domain}). The other four Neumann
lines which form the boundary of $\stardom$ are obtained by noting
that $\stardom$ is symmetric with respect to horizontal and vertical
reflections (see Figure \ref{Fig:quarter_of_star}). Plugging the
expression of the eigenfunction (\ref{eq:Torus-eigenfunction}) in
the flow equations \eqref{eq:flow} we get 
\[
\begin{pmatrix}\dot{x}\\
\dot{y}
\end{pmatrix}=-\frac{\pi}{2}\begin{pmatrix}a^{-1}\cos\left(\frac{\pi}{2a}x\right)\cos\left(\frac{\pi}{2b}y\right)\\
b^{-1}\sin\left(\frac{\pi}{2a}x\right)\sin\left(\frac{\pi}{2b}y\right)
\end{pmatrix}.
\]
Hence, the tangent to any gradient flow line is 
\[
\frac{\ud y}{\ud x}=\frac{a}{b}\tan\left(\frac{\pi}{2a}x\right)\tan\left(\frac{\pi}{2b}y\right).
\]
Integrating this we obtain the gradient flow lines 
\begin{equation}
y(x)=\frac{2b}{\pi}\arcsin\left(\sin\left(\frac{\pi}{2b}y_{0}\right)\left[\cos\left(\frac{\pi}{2a}x\right)\right]^{\left(\frac{a}{b}\right){}^{2}}\right),\label{eq:gradient_flow_line}
\end{equation}
 where $(0,y_{0})$ is a point through which the gradient flow line
passes. Note that for $-b{\color{blue}<}y_{0}{\color{blue}<}b$, each
of the gradient flow lines in (\ref{eq:gradient_flow_line}) is connected
to the extremal point $(a,0)$, but only the one with $y_{0}=b$ is
connected to the saddle point $(0,b)$ and hence it is the desired
Neumann line\footnote{As a matter of fact, $y_{0}=-b$ also gives a Neumann line, but it
is connected to a different saddle point.}.
\end{proof}
\begin{rem}
From the proof of Lemma \ref{lem:gamma_formula} one may also obtain
that there is no gradient flow line which connects two saddle points
of the eigenfunction $f_{a,b}$. From this we conclude that $f_{a,b}$
is a Morse-Smale function \cite[Proposition A.7]{AloBanBerEgg_Neumann}.
\end{rem}

The next lemma shows that the boundary of $\stardom$ is regular enough
for applying an appropriate Sobolev space analysis. The classification
of the boundary in the Lemma is based on \cite[Definition~4.1]{Edmunds:1987}.
\begin{lem}
\label{lem:classC} The boundary of the star-like domain, $\partial\stardom$,
is of class $C$.

Namely, for any $\boldsymbol{p}\in\partial\stardom$ there exists
an open neighborhood $U(\boldsymbol{p})\subset\R^{2}$ and a continuous
function $h\in C(I)$ on an interval $I\subset\rz$ such that for
suitable local Cartesian coordinates
\begin{equation}
\partial\stardom\cap U(\boldsymbol{p})=\left\{ (s,t):\ t=h(s),\ s\in I\right\} \label{eq:class_C_condition}
\end{equation}
holds.
\end{lem}

\begin{proof}
Since the boundary consists of gradient flow lines the claim is obvious
for every point $\boldsymbol{p}$ not being an end point of such a
flow line (i.e, for every $\bs p$ which is not a critical point).
At a saddle point any two adjacent Neumann lines meet with an angle
of $\tfrac{\pi}{2}$, \cite[Theorem~3.2.]{McDFul_ptrs13}. Hence,
the boundary at a neighborhood of a saddle point is also a continuous
function. At the extremal points $(\pm a,0)$ adjacent Neumann lines
meet with an angle of $0$ and form a cusp. We derive the asymptotics
of $\gamma_{a,b}(a-x)$, $x\rightarrow0^{+}$. Using
\begin{align*}
\cos\left(\frac{\pi}{2a}(a-x)\right) & =\sin\left(\tfrac{\pi}{2a}x\right)=\tfrac{\pi}{2a}x+\Or\left(\left(\tfrac{\pi}{2a}x\right){}^{3}\right)\\
(1+x)^{\beta} & =1+\Or(x)\quad\quad\textrm{for }\beta>0\\
\arcsin(x) & =x+\Or\left(x^{3}\right),
\end{align*}
we get that for $x\rightarrow0^{+}$

\begin{equation}
\gamma_{a,b}(a-x)=\frac{2b}{\pi}\arcsin\left(\left[\cos\left(\frac{\pi}{2a}(a-x)\right)\right]^{\left(\frac{a}{b}\right)^{2}}\right)=\frac{2b}{\pi}\left(\frac{\pi}{2a}x\right){}^{\left(\frac{a}{b}\right){}^{2}}+\Or\left(x^{3\left(\frac{a}{b}\right){}^{2}}\right).\label{asymp}
\end{equation}

These asymptotics show that $\gamma_{a,b}$ is strictly monotonically
decreasing in a left neighborhood of $(a,0)$ and its inverse exists
there. Hence, the condition (\ref{eq:class_C_condition}) is satisfied
in a neighborhood of $(a,0)$ by choosing
\[
h(s)=\begin{cases}
\gamma_{a,b}^{-1}(s) & s>0\\
\gamma_{a,b}^{-1}(-s) & s<0
\end{cases}.
\]
\end{proof}
Finally. we use the expression of $\gamma_{a,b}$ to bound the area
of $\stardom$ which is needed in the proofs of Theorem \ref{thm:Spectral-pos-separable},(\ref{enu:thm-spectral-pos-lens})
(first proof) and Lemma \ref{lem:eigenvalue-inequality-vert-sect}
(see (\ref{eq:bound_on_star_area}) in that proof).
\begin{lem}
\label{lem:bound_on_star_area} There exists $c>1$ such if $\nicefrac{a}{b}>c$
then 
\begin{equation}
\frac{1}{ab}\left(\frac{b}{a}+\frac{a}{b}\right)\left|\stardom\right|<\frac{2}{\pi}(j_{0})^{2}\label{eq:Lemma-bound_on_star_area}
\end{equation}
where $j_{0}\approx2.4048$ is the first zero of $J_{0}$, the zeroth
Bessel function.
\end{lem}

\begin{figure}[H]
\centering{}\includegraphics[width=1\textwidth]{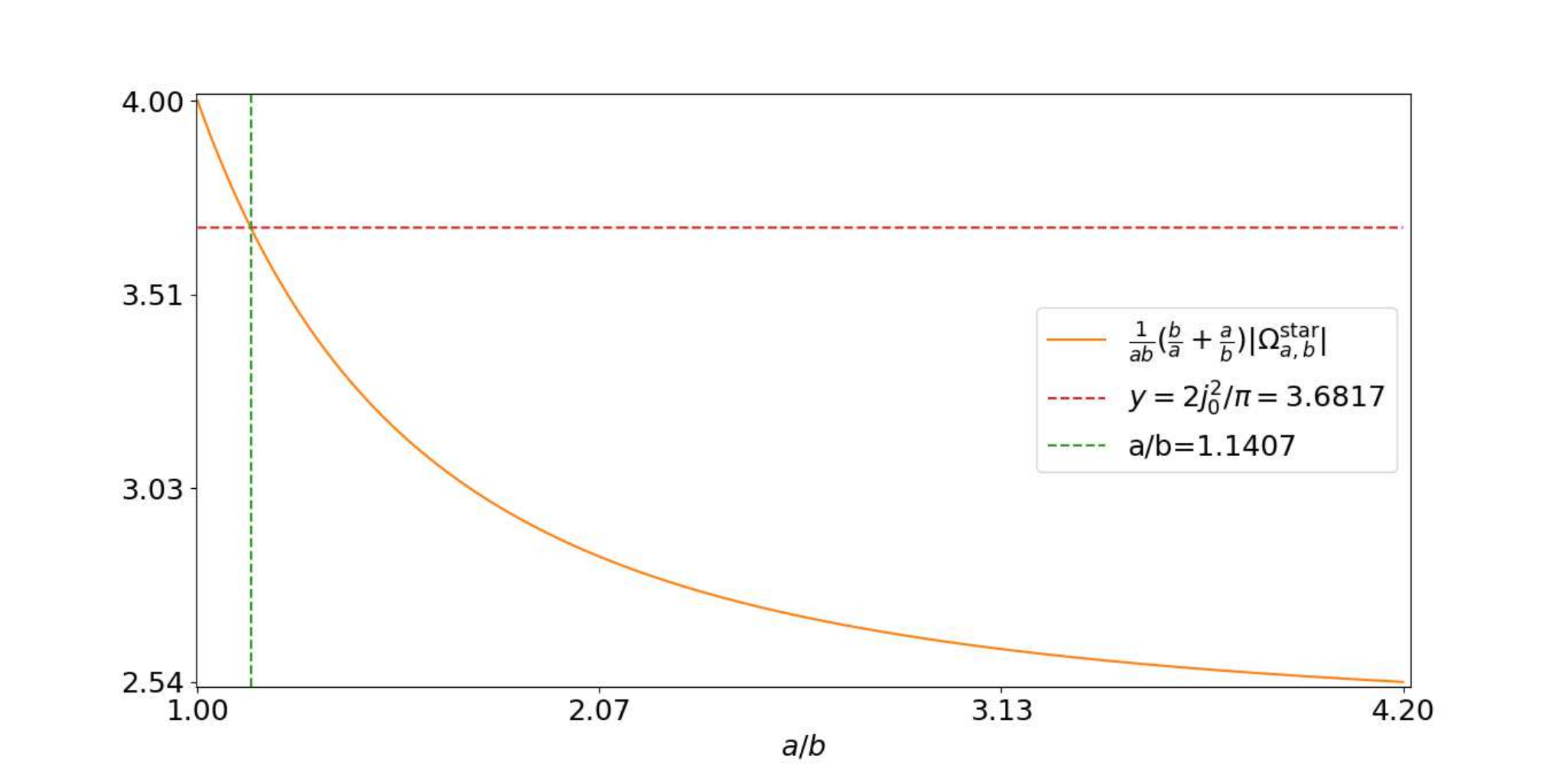} \caption{The left hand side of (\ref{eq:Lemma-bound_on_star_area}) plotted
as a function of $\nicefrac{a}{b}$. The right hand side of (\ref{eq:Lemma-bound_on_star_area})
is indicated together with the corresponding $\nicefrac{a}{b}$ value.}
\label{Fig:Area_of_star} 
\end{figure}

\begin{proof}
Using Lemma \ref{lem:gamma_formula} we have 
\begin{align}
\frac{1}{ab}\left|\stardom\right| & =\frac{8}{\pi a}\int_{0}^{a}\arcsin\left(\left[\cos\left(\frac{\pi}{2a}x\right)\right]^{\left(\frac{a}{b}\right)^{2}}\right)\ud x\nonumber \\
 & =\frac{16}{\pi^{2}}\int_{0}^{\nicefrac{\pi}{2}}\arcsin\left(\left[\cos\left(z\right)\right]^{\left(\frac{a}{b}\right)^{2}}\right)\ud z.\label{eq:area_as_integral}
\end{align}
We may use the Taylor expansion of $\mathrm{ln}\left[\cos\left(z\right)\right]$,
which converges for $\left|z\right|<\frac{\pi}{2}$ (see e.g., \cite[4.3.72]{AbrSte_book64}
and \cite[p. 27]{MagObeSon_book66}) to obtain the bound 
\begin{equation}
\forall z\in(0,\frac{\pi}{2}),\quad\left[\cos\left(z\right)\right]^{\left(\frac{a}{b}\right)^{2}}<\exp\left[-\frac{1}{2}\left(\frac{a}{b}\right)^{2}z^{2}\right].\label{eq:bound_on_cos}
\end{equation}

Another bound which we use is 
\begin{equation}
\forall w\in(0,1),\quad\arcsin\left(w\right)\leq w+\left(\nicefrac{\pi}{2}-1\right)w^{3}.\label{eq:bound_on_arcsin}
\end{equation}
To validate (\ref{eq:bound_on_arcsin}) we may observe that both functions
at the RHS and LHS coincide for $w=0$ and $w=1$ and further check
that the difference does not vanish anywhere else in $\left(0,1\right)$
(for example, since the difference has only a single critical point
in this interval).

Plugging the bounds (\ref{eq:bound_on_cos}), (\ref{eq:bound_on_arcsin})
in (\ref{eq:area_as_integral}) and using also the monotonicity of
$\arcsin(w)$ for $w\in(0,1)$ we get 
\begin{align}
\frac{1}{ab}\left|\stardom\right| & <\frac{16}{\pi^{2}}\int_{0}^{\nicefrac{\pi}{2}}\left\{ \exp\left[-\frac{1}{2}\left(\frac{a}{b}\right)^{2}z^{2}\right]+\left(\nicefrac{\pi}{2}-1\right)\exp\left[-\frac{3}{2}\left(\frac{a}{b}\right)^{2}z^{2}\right]\right\} \ud z\nonumber \\
 & <\frac{16}{\pi^{2}}\int_{0}^{\infty}\left\{ \exp\left[-\frac{1}{2}\left(\frac{a}{b}\right)^{2}z^{2}\right]+\left(\nicefrac{\pi}{2}-1\right)\exp\left[-\frac{3}{2}\left(\frac{a}{b}\right)^{2}z^{2}\right]\right\} \ud z\nonumber \\
 & =\frac{16}{\pi^{2}}\left\{ \frac{1}{2}\sqrt{\frac{2\pi}{\left(\nicefrac{a}{b}\right)^{2}}}+\left(\nicefrac{\pi}{2}-1\right)\frac{1}{2}\sqrt{\frac{2\pi}{3\left(\nicefrac{a}{b}\right)^{2}}}\right\} \approx2.7014\cdot\frac{b}{a},\label{eq:bound-on-star-area}
\end{align}
where moving to the last line we used integration over (half) Gaussian.

From the above we get $\frac{1}{ab}\left(\frac{a}{b}+\frac{b}{a}\right)\left|\stardom\right|\apprle2.7014\cdot\left(1+\left(\frac{b}{a}\right)^{2}\right)$.
Now, since $\frac{2}{\pi}(j_{0})^{2}\approx3.68$ we get that (\ref{eq:Lemma-bound_on_star_area})
holds if $\frac{b}{a}$ is small enough.
\end{proof}
\begin{rem}
\label{rem:Numeric_value_of_constant}From the proof one easily gets
that (\ref{eq:Lemma-bound_on_star_area}) holds for $\nicefrac{a}{b}\gtrsim\left(\frac{3.68}{2.7}-1\right)^{-\nicefrac{1}{2}}\approx1.66$.
Numerically, it seems that choosing $c\approx1.1407$ already guarantees
this bound. This can be seen in Figure \ref{Fig:Area_of_star}.
\end{rem}

{\small{}\bibliographystyle{abbrv}
\bibliography{GlobalBib_190822}
 } 
\end{document}